\documentstyle[aps,eqsecnum]{revtex}
\begin{document}
\draft
\title{Monte Carlo evaluation of path integrals for the nuclear shell
model}
\author{G.~H.~Lang, C.~W.~Johnson\thanks{Present Address: Los Alamos
National Laboratory, T-5, Mail Stop B283, P.~O.~Box 1663, Los Alamos,
NM 87545.}, S.~E.~Koonin, and W.~E.~Ormand}
\address{W.~K.~Kellogg Radiation Laboratory, 106-38\\
California Institute of Technology, Pasadena, CA 91125}
\maketitle
\begin{abstract}
We present in detail a formulation of the shell model as a path
integral and Monte Carlo techniques for its evaluation. The
formulation, which linearizes the two-body interaction by an
auxiliary field, is quite general, both in the form of the effective
`one-body' Hamiltonian and in the choice of ensemble. In particular,
we derive formulas for the use of general (beyond monopole) pairing
operators, as well as a novel extraction of the canonical
(fixed-particle number) ensemble via an activity expansion. We
discuss the advantages and disadvantages of the various formulations
and ensembles and give several illustrative examples. We also discuss
and illustrate calculation of the imaginary-time response function
and the extraction, by maximum entropy methods, of the corresponding
strength function. Finally, we discuss the ``sign-problem'' generic
to fermion Monte Carlo calculations, and prove that a wide class of
interactions are free of this limitation.
\end{abstract}


\parskip=6pt
\section{Motivation and Organization}

Exact diagonalizations of shell model Hamiltonians in the $0s$-$1d$
shell demonstrated \cite{wild84} that the shell model can yield an
accurate and consistent description for a wide range of nuclear
properties in different nuclei, if the many-body basis is
sufficiently large. However, the combinatorial scaling of the
many-body space with the size of the single particle basis or the
number of valence nucleons restricts such exact diagonalizations to
small nuclei or nuclei with few valence particles \cite{wild88}.

In facing the general challenge of developing non-perturbative
methods to describe strongly interacting many-body systems, various
quantum Monte Carlo schemes have been proposed as an alternative to
direct diagonalization \cite{linden92}. Among them, the
auxiliary-field Monte Carlo method is suitable for addressing
interacting fermions \cite{koon82}. This method is a Monte Carlo
evaluation of the path-integral obtained by the Hubbard-Stratonovich
transformation \cite{hub59} of the imaginary-time evolution operator.
The many-body wavefunction is represented by a set of determinantal
wavefunctions evolving in fluctuating auxiliary fields. The method
thus enforces the Pauli principle exactly, and the storage and
computation time scale gently with the single-particle basis or the
number of particles. Auxiliary field methods have been applied to
condensed matter systems such as the Hubbard
Model\cite{linden92,hirsch83}, yielding important information about
electron correlations and magnetic properties.

In this paper, we discuss the application of auxiliary-field Monte
Carlo techniques to the nuclear shell model. This involves a two-body
interaction more general than the simple on-site repulsion of a
Hubbard model. Our goal is to develop new methods for extending the
applicability of the shell model, as well as to investigate the
powers and limitations of auxiliary-field Monte Carlo methods for
general fermion systems.

We have previously published a letter \cite{johnson92} with selected
results for static observables in {\it sd}- and {\it fp}-shell
nuclei. This paper serves to give the details of the implementation,
introduce a method for calculating dynamical correlations and
strength functions, demonstrate the method with simple ({\it
sd}-shell) nuclei, and discuss several important issues that arise in
the implementation. We also explore the limitations imposed by the
negative contributions in the path-integral, referred to as the
``sign problem'' in the literature.

Our presentation is organized as follows. We begin in Section II by
using the Hubbard-Stratonovich (HS) transformation to write the
imaginary time evolution operator $\exp(-\beta \hat{H})$ as a path
integral. This requires that the Hamiltonian $\hat{H}$ be cast as a
quadratic form using an appropriate set of operators. We discuss in
Section~III two ways in which this can be accomplished, using either
particle density or pairing operators. The imaginary time evolution
operator can be used to extract information about the system at
finite temperature or in its ground state ($\beta \rightarrow
\infty$). The formulae for obtaining static observables are presented
in Section~IV where methods for handling the canonical ensemble are
also introduced. We then discuss in Section~V the extraction of the
strength functions for operators from the imaginary time response
function. In Section~VI, we briefly describe the computational
algorithms for implementing our methods and present selected results
of calculations for {\it sd}-shell nuclei. In the final section, we
address the sign problem and also discuss a class of non-trivial
interactions that give rise to a positive definite path-integral for
some nuclei.

\section{The imaginary time evolution operator}

Given some many-body Hamiltonian $\hat{H}$, we seek a tractable
expression for the imaginary time evolution operator:
\begin{equation}
\hat{U} = \exp ( -\beta \hat{H})\;.
\end{equation}
Here, $\beta$ has units of inverse energy and $\beta^{-1}$ can be
interpreted as an imaginary time. (Here and throughout, we take
$\hbar=1$ and measure all energies in units of MeV.) It is also clear
that $\hat{U}$ can be interpreted as the partition operator for
temperature $\beta^{-1}$. We will refer to $\hat{U}$ as the evolution
operator hereafter. The operator $\hat{H}$ is usually a generalized
Hamiltonian and might contain terms beyond the true Hamiltonian, such
as $-\mu \hat{N}$ in the grand-canonical ensemble or $-\omega
\hat{J}_z$ if we are `cranking' the system.

There are two formalisms for extracting information from the
evolution operator: the ``thermal'' formalism (on which we will
concentrate) and the ``zero-temperature'' formalism (to which the
thermal formalism reduces in the limit $\beta \rightarrow \infty$).
In the thermal formalism, we begin with the partition function
\begin{equation}
Z = {\rm \hat{T}r} \exp ( -\beta \hat{H})\;,
\end{equation}
and then construct the thermal observable of an operator $\hat{O}$:
\begin{equation}
\left\langle \hat{O} \right\rangle =
{1 \over Z} {\rm \hat{T}r} \left[ \hat{O} \exp ( -\beta \hat{H})
\right]\;.
\end{equation}
Here, the trace ${\rm \hat{Tr}}$ is over many-body states of fixed
(canonical) or all (grand-canonical) particle number. In the
zero-temperature formalism we begin with a trial wavefunction
$\psi_0$ and use the evolution operator to project out the ground
state, assuming that $\psi_0$ is not orthogonal to the ground state.
The expectation value of $\hat{O}$ is then given by
\begin{equation}
\left\langle \hat{O} \right\rangle = \lim_{\beta \rightarrow
\infty}
{{\left\langle \psi_0 \left|
\exp ( -{\beta \over 2} \hat{H})
\hat{O} \exp ( -{\beta\over 2} \hat{H})
\right| \psi_0 \right\rangle}
\over
{\left\langle \psi_0 \left|
\exp ( -\beta \hat{H})
\right| \psi_0 \right\rangle}}\;.
\end{equation}
In this section, we describe how to write $\hat{U}$ in a form that
allows Eqs.~(2.3) or (2.4) to be evaluated.

\subsection{Path integral formulation of the evolution operator}

We restrict ourselves to generalized Hamiltonians that contain at
most two-body terms. The Hamiltonian $\hat{H}$ can then be written as
a quadratic form in some set of `convenient' operators
$\hat{\cal{O}}_\alpha $:
\begin{equation}
\hat{H} = \sum_{\alpha} \epsilon_{\alpha} \hat{\cal{O}}_{\alpha}
+ {1 \over 2} \sum_{\alpha} V_{\alpha} \hat{\cal{O}}_{\alpha}^2\;,
\end{equation}
where we've assumed that the quadratic term is diagonal in the
$\hat{\cal{O}}_{\alpha}$. The meaning of `convenient' will become
clear shortly, but typically it refers to one-`body' operators,
either one-particle (`density') or one-quasiparticle (`pairing'). The
strength of the two-body interaction is characterized by the real
numbers $V_{\alpha}$.

For $\hat{H}$ in the quadratic form (2.5), one can write the
evolution operator $\hat{U}$ as a path integral. The exponential is
first split into $N_t$ `time' slices, $\beta = N_t \Delta\beta$, so
that
\begin{equation}
\hat{U} = \left[ \exp ( -\Delta\beta\hat{H})\right]^{N_t}\;.
\end{equation}
Then we perform the Hubbard-Stratonovich (HS) transformation on the
two-body term for the $n$'th time slice to give \cite{hub59},
\begin{equation}
\exp \left(-\Delta \beta \hat{H}\right)\simeq
\int_{-\infty}^{\infty} \prod_{\alpha}d\sigma_{\alpha n}
\left({\Delta \beta |V_\alpha|\over{2\pi}}\right)^{1/2}
\exp \left\{-\Delta\beta
\left(\sum_{\alpha}{1\over2}|V_\alpha|\sigma_{\alpha n}^2
+\epsilon_\alpha
\hat{\cal O}_{\alpha}+s_\alpha V_\alpha \sigma_{\alpha n}
\hat{\cal O}_\alpha\right)\right\}\;,
\end{equation}
where the phase factor $s_\alpha$ is $\pm 1$ if $V_{\alpha} < 0$ and
is $\pm i$ if $V_{\alpha} > 0$. Each real variable $\sigma_{\alpha
n}$ is the {\it auxiliary field} associated with $\hat{\cal
O}_\alpha$ at time slice $n$.

The approximation (2.7) is valid through order $\Delta \beta$, since
the corrections are commutator terms of order $(\Delta \beta)^2$. The
evolution operator is then
\begin{equation}
\hat{U}=\left[\exp \left(-\Delta \beta
\hat{H}\right)\right]^{N_t}\simeq
\int {\cal D}^{N_t}[\sigma]
G(\sigma)\exp \left(-\Delta \beta h_{\sigma}\left
(\tau_{N_t}\right)\right)\ldots\exp \left(-\Delta \beta
h_\sigma \left(\tau_1 \right) \right)
\end{equation}
where the integration measure is
\begin{mathletters}
\begin{equation}
{\cal D}^{N_t}[\sigma]=\prod_{n=1}^{N_t} \prod_{\alpha}
d\sigma_{\alpha n}
\left(\Delta\beta |V_{\alpha}|\over2\pi\right)^{1\over 2}\;,
\end{equation}
the Gaussian factor is
\begin{equation}
G(\sigma)=\exp\left(-\sum_{\alpha n} {1 \over 2}|V_{\alpha}|
\sigma_{\alpha n}^2\right)\;,
\end{equation}
\end{mathletters}
and the one-body hamiltonian is
\begin{equation}
\hat{h}_{\sigma} (\tau_n) =
\sum_{\alpha} \left( \epsilon_\alpha + s_\alpha V_\alpha
\sigma_{\alpha n} \right) \hat{\cal{O}}_\alpha\;.
\end{equation}

It is sometimes convenient to employ a continuum notation,
\begin{eqnarray}
\hat{U} &=& \int {\cal D} \left[\sigma \right]
\exp \left( -{1\over 2} \int_0^{\beta} d\tau \,
\sum_{\alpha} \left| V_{\alpha} \right|
\sigma_{\alpha}^2(\tau)\right)\nonumber\\
&&\times \left[
{\cal T} \exp \left( -\int_0^{\beta} d\tau \, \hat{h}_{\sigma}(\tau)
\right) \right]\;,
\end{eqnarray}
where ${\cal T}$ denotes time-ordering and
\begin{equation}
{\cal D}\left[ \sigma \right] =\lim_{N_t\rightarrow \infty}
{\cal D}^{N_t}\left[ \sigma \right]\;,
\end{equation}
\begin{equation}
{\cal T} \exp \left( -\int_0^{\beta} d\tau \,
\hat{h}_{\sigma}(\tau) \right) =
\lim_{N_t \rightarrow \infty}
\prod_{n=1}^{N_t} \exp \left( -\Delta \beta \,
\hat{h}_\sigma(\tau_n) \right)\;.
\end{equation}

In the limit of an infinite number of time slices Eq.~(2.8) is exact.
In practice one has a finite number of time slices and the
approximation is valid only to order $\Delta \beta$. The case of only
one time slice is known as the Static Path Approximation (SPA);
previous work on the SPA and its extensions can be found in
Refs.~\cite{arve88} and \cite{pudd91}.

Rewriting the evolution operator as a path integral can make the
model space tractable. Consider the case where the
$\hat{\cal{O}}_\alpha$ are density operators. Then Eq.~(2.1) is an
exponential of two-body operators; it acts on a Slater-determinant to
produce a sum of many Slater-determinants. In contrast, the
path-integral formulation (2.8) contains only exponentials of
one-body operators which, by Thouless' theorem \cite{thouless}, takes
a Slater-determinant to another single Slater-determinant. Therefore,
instead of having to keep track of a very large number of
determinants (often many thousands for modern matrix-diagonalization
shell model codes such as OXBASH \cite{brown85}), we need deal only
with one Slater-determinant at a time. Of course, the price to be
paid is the evaluation of a high-dimensional integral. However, the
number of auxiliary fields grows only quadratically with the size of
the single particle basis while the corresponding number of
Slater-determinants grows exponentially. Furthermore, the integral
can be evaluated stochastically, making the problem ideal for
parallel computation.

\subsection{Monte Carlo evaluation of the path integral}

Formulating the evolution operator as a path integral over auxiliary
fields reduces the problem to quadrature. For a limited number of
auxiliary fields, such as in the SPA with only a
quadrupole-quadrupole interaction, the integral can be evaluated by
direct numerical quadrature. However, for more general cases
(typically hundreds of fields), the integral must be evaluated
stochastically using Monte Carlo techniques.

Using the one-body evolution operator defined by
\begin{equation}
\hat{U}_\sigma(\tau_2,\tau_1) = {\cal{T}}
\exp \left(-\int_{\tau_1}^{\tau_2}
d\tau \, \hat{h}_\sigma(\tau)\right)\;,
\end{equation}
we can write Eq.~(2.3) or (2.4) as
\begin{equation}
\left\langle \hat{O} \right\rangle =
{{
\int {\cal D}\left[\sigma \right] \,
G(\sigma) \left\langle \hat{O}(\sigma) \right\rangle
\zeta(\sigma)}
\over
{\int {\cal D}\left[\sigma \right] \,
G(\sigma)\, \zeta(\sigma)}}\;.
\end{equation}
For the zero-temperature formalism
\begin{equation}
\zeta(\sigma) \equiv
\left\langle \psi_0 \left|
\hat{U}_{\sigma}(\beta,0)
\right| \psi_0 \right\rangle
\end{equation}
and
\begin{equation}
\left\langle \hat{O} (\sigma)\right\rangle =
{{\left\langle \psi_0 \left|
\hat{U}_{\sigma}(\beta,\beta/2) \,
\hat{O} \,
\hat{U}_{\sigma}(\beta/2,0)
\right| \psi_0 \right\rangle}
\over
{\left\langle \psi_0 \left|
\hat{U}_{\sigma}(\beta,0)
\right| \psi_0 \right\rangle}}\;,
\end{equation}
while for the thermal formalism (canonical and grand-canonical),
\begin{equation}\zeta(\sigma)\equiv
{\rm\hat{T}r}[\hat{U}_\sigma(\beta,0)]\;,
\end{equation}
and
\begin{equation}
\left\langle \hat{O} \right\rangle_{\sigma}
= {{{\rm \hat{T}r} \left[ \hat{O}\, \hat{U}_{\sigma}(\beta,0)
\right]}
\over{ {\rm \hat{T}r}\hat{U}_{\sigma}(\beta,0)}}\;.
\end{equation}

To evaluate the path integral via Monte Carlo techniques, we must
choose a normalizable positive-definite weight function $W_{\sigma}$,
and generate an ensemble of statistically independent fields $\left\{
\sigma_i \right\}$ such that the probability density to find a field
with values $\sigma_i$ is $W_{\sigma_i}$. Defining the `action' by
\begin{equation}
{\cal S}_\sigma=\sum_\alpha{1\over 2}
|V_\alpha|\int_0^{\beta}
d\tau \sigma_\alpha(\tau)^2-\ln \zeta(\sigma)\;,
\end{equation}
the required observable is then simply
\begin{equation}
\left\langle \hat{O} \right\rangle =
{{\int {\cal D}\left[ \sigma \right]
\left\langle \hat{O} \right\rangle_{\sigma}
e^{-{\cal S}_{\sigma}}}
\over
{\int {\cal D}\left[ \sigma \right] e^{-{\cal S}_\sigma}}}=
{{{1 \over N} \sum_i
\left\langle \hat{O} \right\rangle_{i} \Phi_i}
\over
{{1\over N}\sum_i \Phi_i}}\;,
\end{equation}
where $N$ is the number of samples,
\begin{equation}
\Phi_i=e^{-{\cal S}_i}/W_i
\end{equation}
and ${\cal S}_i \equiv {\cal S}_{\sigma_i}$, etc. Ideally $W$ should
approximate $\exp(-{\cal S})$ closely. However, $\exp(-{\cal S})$ is
generally not positive and can even be complex. In some cases,
$\Phi_i$ may oscillate violently, giving rise to a very small
denominator in Eq.~(2.21) to be cancelled by a very small numerator.
While this cancellation is exact analytically, it is only approximate
in the Monte Carlo evaluation so that this `sign problem' leads to
large variances in the evaluation of the observable.

There are several possible schemes for both the choice of $W$ and the
sampling of the fields. We typically choose $W= \left| \exp(-{\cal
S}) \right|$ and generate the samples via random walk (Metropolis)
methods.

\section{Decompositions of the Hamiltonian}

To realize the HS transformation, the two-body parts of $\hat{H}$
must be cast as a quadratic form in one-body operators
$\hat{\cal{O}}_{\alpha}$. As these latter can be either density
operators or pair creation and annihilation operators (or both),
there is considerable freedom in doing so. In the simplest example,
let us consider an individual interaction term,
\begin{equation}
\hat{H}=a_{1}^{\dagger} a_{2}^{\dagger} a_4 a_3\;,
\end{equation}
where $a_{i}^{\dagger}, a_{i}$ are anti-commuting fermion creation
and annihilation operators. In the pairing decomposition, we write
(using the upper and lower bracket to indicate the grouping)
\begin{mathletters}
\begin{eqnarray}
\hat{H} & = & \overbrace{a_1^{\dagger}
a_2^{\dagger}} \underbrace{a_4 a_3}\\
& = & {1 \over 4}(a_{1}^{\dagger} a_{2}^{\dagger} +
a_3 a_4)^2-{1\over 4}
(a_{1}^{\dagger}a_{2}^{\dagger} - a_3 a_4)^2+{1 \over2}
[a_{1}^{\dagger} a_{2}^{\dagger},a_3 a_4]\;.
\end{eqnarray}
\end{mathletters}

The commutator is a one-body operator that can be put directly in the
one-body Hamiltonian $\hat{h}_{\sigma}$. The remaining two quadratic
forms in pair-creation and -annihilation operators can be coupled to
auxiliary fields in the HS transformation.

In the density decomposition, there are two ways to proceed: we can
group $(1,3) $ and $(2,4)$ to get
\begin{mathletters}
\begin{eqnarray}
\hat{H} & = & \overbrace{a_1^{\dagger}a_3}
\underbrace{a_2^{\dagger}a_4}
-a_1^{\dagger}a_4 \delta_{23}\\
& = & -a_{1}^{\dagger}a_4 \delta_{23}+
{1 \over 2}[a_{1}^{\dagger}a_3,
a_{2}^{\dagger}a_4]+{1\over4}
(a_{1}^{\dagger}a_3+a_{2}^{\dagger}a_4)^2
-{1\over 4} (a_{1}^{\dagger}a_3 -a_{2}^{\dagger}a_4)^2\;,
\end{eqnarray}
\end{mathletters}
or group $(1,4)$ and $(2,3)$ to get
\begin{mathletters}
\begin{eqnarray}
\hat{H} & = & -\overbrace {a_{1}^{\dagger}a_4}
\underbrace{ a_{2}^{\dagger}
a_3}+a_{1}^{\dagger}a_3 \delta_{24}\;.\\
& = & a_{1}^{\dagger}a_3 \delta_{24} -
{1\over 2} [a_{1}^{\dagger} a_4,
a_{2}^{\dagger}a_3]-{1\over 4}(a_{1}^{\dagger}
a_4+a_{2}^{\dagger}a_3)^{2}
+{1\over 4} (a_{1}^{\dagger} a_4 -
a_{2}^{\dagger} a_3)^2\;.
\end{eqnarray}
\end{mathletters}
Again the commutator terms are one-body operators, but now the
quadratic forms are squares of density operators that conserve
particle number. We refer to Eq.~(3.3) as the `direct' decomposition
and Eq.~(3.4) as the `exchange' decomposition.

For any general two-body Hamiltonian, we can choose the pairing or
density decompositions for different parts of the two-body
interaction. Moreover, even within a pure density break-up
decomposition, there is still freedom to choose between the direct
and exchange formulations. Although the exact path integral result is
independent of the scheme used, different schemes will lead to
different results under certain approximations (e.g.,~mean field or
SPA). The choice of decomposition will also affect the rate of
convergence of our numerical result as $N_t \rightarrow \infty$, as
well as the statistical precision of the Monte Carlo evaluation. Most
significantly, it affects the fluctuation of $\Phi$ in Eq.~(2.21) and
thus determines the stability of the Monte Carlo calculation. (See
Section~VI below.)

In the application of these methods to the nuclear shell model, it is
particularly convenient to use quadratic forms of operators that
respect rotational invariance, isospin symmetry, and the shell
structure of the system. We introduce these in the following
subsections for both the density and pairing decompositions.

\subsection{Density decomposition}

We begin by ignoring explicit isospin labels and by writing the
rotationally invariant two-body Hamiltonian as
\begin{eqnarray}
\hat{H}_{2} & = & {1 \over 2} \sum_{abcd}\sum_J
V_{J}(ab,cd)\sum_{M} A_{JM}^{\dagger}
(ab)A_{JM}(cd)\nonumber\\
& = & {1 \over 4} \sum_{abcd}\sum_J
\left[ (1+\delta_{ab})(1+\delta_{cd}) \right]^{1/2}
V_{J}^{A}(ab,cd)\sum_{M} A_{JM}^{\dagger}
(ab)A_{JM}(cd)
\end{eqnarray}
where the sum is taken over all proton and neutron single-particle
orbits (denoted by $a,b,c,d$) and the pair creation and annihilation
operators are given by
\begin{mathletters}
\begin{eqnarray}
A_{JM}^{\dagger}(a b) & = &
\sum_{m_{a} m_{b}}(j_a m_a j_b m_b
| J M) a_{j_b m_b}^{\dagger}
a_{j_a m_a}^{\dagger} = -[ a_{j_a}^{\dagger} \times
a_{j_b}^{\dagger} ]^{JM}\\
A_{JM}(ab) & = &
\sum_{m_{a}m_{b}}(j_a m_a j_b m_b| J M) a_{j_a m_a}
a_{j_b m_b} = [ a_{j_a} \times a_{j_b} ]^{JM}\;.
\end{eqnarray}
\end{mathletters}
The $V_J(ab,cd)$ are the angular-momentum coupled two-body matrix
elements of a scalar potential $V(\vec{r}_1,\vec{r}_2)$ defined as
\begin{equation}
V_J(ab,cd) = \langle [\psi_{j_a}(\vec{r}_1)\times
\psi_{j_b}(\vec{r}_2)]^{JM} |V(\vec{r}_1,\vec{r}_2) |
[\psi_{j_c}(\vec{r}_1)\times \psi_{j_d}(\vec{r}_2)]^{JM} \rangle\;,
\end{equation}
(independent of M) while the anti-symmetrized two-body matrix
elements $V_J^A(ab,cd)$ are given by
\begin{equation}
V_J^A(ab,cd)= \left[ (1+\delta_{ab})
(1+\delta_{cd})\right]^{-1/2}
\left[ V_J(ab,cd) -(-1)^{j_c+j_d-J}V_J(ab,dc)\right]\;.
\end{equation}

Before continuing discussion of the density decomposition, we note
that the two-body Hamiltonian for fermion systems is completely
specified by the set of anti-symmetrized two-body matrix elements
$V_J^A(ab,cd)$ that are the input to many standard shell model codes
such as OXBASH \cite{brown85}. Indeed, we can add to the
$V_J^A(ab,cd)$ any set of (unphysical) symmetric two-body matrix
elements $V_J^S(ab,cd)$ satisfying
\begin{equation}
V_J^S(ab,cd) = (-1)^{j_c+j_d-J}V_J^S(ab,dc)\;,
\end{equation}
without altering the action of $\hat{H}_2$ on any many-fermion wave
function. However, note that although the $V_J^S(ab,cd)$ do not alter
the eigenstates and eigenvalues of the full Hamiltonian, they can
(and do) affect the character of the decomposition of $\hat{H}_2$
into density operators, as is shown below. In what follows, we define
the set of two-body matrix elements $V_J^N(ab,cd)$ that may possess
no definite symmetries as
\begin{equation}
V_J^N(ab,cd) = V_J^A(ab,cd)+V_J^S(ab,cd)\;,
\end{equation}
allowing us to write the two-body Hamiltonian as
\begin{equation}
\hat{H}_{2}={1 \over 4} \sum_{abcd}\sum_J
\left[ (1+\delta_{ab})(1+\delta_{cd}) \right
]^{1/2}V_{J}^{N}(ab,cd)\sum_{M}
A_{JM}^{\dagger}(ab)A_{JM}(cd)\;.
\end{equation}

To obtain the density decomposition of $\hat{H}_2$, we perform a
Pandya transformation to recouple $(a,c)$ and $(b,d)$ into density
operators with definite multipolarity
\begin{equation}
\hat{\rho}_{KM}(ab)=\sum_{m_a,m_b}
(j_a m_aj_b m_b|K M)
a^{\dagger}_{j_a m_a}
\tilde{a}_{j_b m_b}\;,
\end{equation}
where $\tilde{a}_{j_a m_a} = (-1)^{j_a+m_a}
a_{j_a m_a}$.
Then $\hat{H}_2$ can be rewritten as
\begin{equation}
\hat{H}_2=\hat{H}'_{2}+\hat{H}'_{1}\;;
\end{equation}
\begin{equation}
\hat{H}_{2}'={1 \over 2}\sum_{abcd}\sum_K E_K
(ac,bd)\sum_M (-1)^{M}
\hat{\rho}_{K-M}(ac)\hat{\rho}_{KM}(bd)\;,
\end{equation}
where the particle-hole matrix elements of the interaction are
\begin{eqnarray}
E_{K}(ac,bd) &=& (-1)^{j_b+j_c} \sum_J (-1)^J (2J+1)
\left\{
\matrix{ j_a & j_b & J \cr
j_d & j_c & K \cr}
\right\}\nonumber\\
&&\times
{1\over 2} V^{N}_{J}(ab,cd) \sqrt{(1+\delta_{ab}) (1+\delta_{cd})}\;,
\end{eqnarray}
and $\hat{H}_{1}'$ is a one-body operator given by
\begin{mathletters}
\begin{equation}
\hat{H}'_1 = \sum_{ad}
\epsilon'_{ad}\hat{\rho}_{0\, 0}(a,d)\;,
\end{equation}
with
\begin{equation}
\epsilon'_{ad} = -{1\over 4} \sum_b \sum_J
(-1)^{J+j_a+j_b} (2J+1){1 \over \sqrt{2 j_a +1}}
V^{N}_{J}(ab,bd)\sqrt{(1+\delta_{ab}) (1+\delta_{cd})}\;.
\end{equation}
\end{mathletters}
Note that adding symmetric matrix elements is equivalent to using the
exchange decomposition for some parts of the interaction. The freedom
in choosing the combinations of direct and exchange decomposition is
then embodied in the arbitrary symmetric part of the matrix elements
$V^N_J$.

Introducing the shorthand notation $i=(ac),j=(bd)$, we can write
Eq.~(3.14) as
\begin{equation}
\hat{H}'_2 = {1\over 2} \sum_{ij} \sum_{K}
E_{K}(i,j)
(-1)^M
\hat{\rho}_{KM}(i)\hat{\rho}_{K-M}(j)\;.
\end{equation}
Upon diagonalizing the matrix $E_K(i,j)$ to obtain eigenvalues
$\lambda_{K\alpha}$ and associated eigenvectors $v_{K\alpha}$, we can
represent $\hat{H}'_{2}$ as
\begin{equation}
\hat{H}'_2 ={1\over 2}\sum_{K\alpha}\lambda_{K}(\alpha)
(-1)^M \hat{\rho}_{KM}(\alpha) \hat{\rho}_{K-M}(\alpha)\;,
\end{equation}
where
\begin{equation}
\hat{\rho}_{KM}(\alpha)=\sum_i
\hat{\rho}_{KM}(i)v_{K\alpha}(i)\;.
\end{equation}

Finally, if we define
\begin{mathletters}
\begin{equation}
\hat{Q}_{KM}(\alpha)\equiv {1 \over \sqrt{2(1+\delta_{M0})}}
\left(\hat{\rho}_{KM}(\alpha)+(-1)^{M}\hat{\rho}_{K-M}(\alpha)
\right)\;,
\end{equation}
\begin{equation}
\hat{P}_{KM}(\alpha)\equiv -{i\over \sqrt{2(1+\delta_{M0})}}
\left(\hat{\rho}_{KM}(\alpha)-(-1)^{M}\hat{\rho}_{K-M}(\alpha)
\right)\;,
\end{equation}
\end{mathletters}
then $\hat{H}_{2}'$ becomes
\begin{equation}
\hat{H}'_2={1 \over 2}\sum_{K \alpha}\lambda_K
(\alpha)\sum_{M\geq 0}
\left(\hat{Q}_{KM}^{2}(\alpha)+\hat{P}_{KM}^{2}
(\alpha)\right)\;.
\end{equation}
This completes the representation of the two-body interaction as a
diagonal quadratic form in density operators. We then couple
auxiliary fields $\sigma_{KM}(\alpha)$ to $\hat{Q}_{KM}$ and
$\tau_{KM}(\alpha)$ to $\hat{P}_{KM}$ in the HS transformation. (The
latter are not to be confused with the ``imaginary time'' $\tau$.)

In the treatment thus far, protons and neutrons were not
distinguished from each other. Although the original Hamiltonian
$\hat{H}_2$ conserves proton and neutron numbers, we ultimately might
deal with one-body operators $\rho_{KM}(a_p,b_n)$ and
$\rho_{KM}(a_n,b_p)$ ({\it n,p} subscripts denoting neutron and
proton) that individually do not do so. The one-body hamiltonian
$\hat{h}_\sigma$ appearing in the HS transformation then mixes
neutrons and protons. The single-particle wavefunctions in a Slater
determinant then contain both neutron and proton components and
neutron and proton numbers are not conserved separately in each Monte
Carlo sample; rather the conservation is enforced only statistically.

It is, of course, possible to recouple so that only density operators
separately conserving neutron and proton numbers
($\hat{\rho}_{KM}(a_p,b_p)$ and $\hat{\rho}_{KM}(a_n,b_n)$) are
present. To do so, we write the two-body Hamiltonian in a manifestly
isospin-invariant form,
\begin{equation}
\hat{H}_2={1\over 4} \sum_{abcd}\sum_J
\left[(1+\delta_{ab})(1+\delta_{cd})\right]^{1/2}
V_{JT}^{N}(ab,cd)A^{\dagger}_{JT;MTz}
(ab)A_{JT;M Tz}(cd)\;,
\end{equation}
where, similar to the previous definition (3.6), the pair operator is
\begin{equation}
A^{\dagger}_{JT;MTz}(ab)=\sum_{m_a,m_b}
(j_a m_a,j_b m_b|JM)({1 \over 2} t_a,{1 \over 2} t_b|T
T_z)a^{\dagger}_{j_b m_b
t_b}a^{\dagger}_{j_a m_a t_a}\;.
\end{equation}
Here $({1 \over 2},t_a)$, etc. are the isospin indices with $t_a=-{1
\over 2}$ for proton states and $t_a={1 \over 2}$ for neutron states,
and $(T T_z)$ are the coupled isospin quantum numbers. The two-body
Hamiltonian can now be written solely in terms of density operators
that conserve the proton and neutron numbers. Namely,
\begin{equation}
\hat{H}_2 = \hat{H}'_1 + \hat{H}'_2\;,
\end{equation}
where
\begin{equation}
\hat{H}'_1 = \sum_{ad} \sum_{t = p,n}
\epsilon'_{ad} \rho_{0\, 0,t}(a,d)\;,
\end{equation}
with
\begin{equation}
\epsilon'_{ad} = -{1\over 4} \sum_b \sum_J
(-1)^{J+j_a+j_b} (2J+1){1 \over \sqrt{2 j_a +1}}
V^{N}_{J,T=1}(ab,bd)\sqrt{(1+\delta_{ab})(1+\delta_{cd})}\;,
\end{equation}
and
\begin{equation}
\hat{H}'_2 = {1\over 2} \sum_{abcd} \sum_{K, T=0,1} E_{KT}(ac,bd)
[\hat{\rho}_{K,T}(i)\times\hat{\rho}_{K,T}(j)]^{J=0}\;.
\end{equation}
Here, we define $\hat{\rho}_{KM,T}$ as
\begin{equation}
\hat{\rho}_{K\,M,T} = \hat{\rho}_{K\, M,p} +(-1)^T
\hat{\rho}_{K\, M,n}\;,
\end{equation}
and the $E_{K,T}$ are given by
\begin{eqnarray}
E_{K,T=0}(ac,bd) & = & (-1)^{j_b+j_c} \sum_J (-1)^J (2J+1)
\left\{ \matrix{ j_a & j_b & J \cr j_d & j_c & K \cr}\right\}
\sqrt{(1+\delta_{ab})(1+\delta_{cd})}\nonumber\\
& & \times
{1 \over 2} \left[
V^N_{J,T=1}(ab,cd) +{1\over
2}(V_{J,T=0}^{A}(ab,cd)-V_{J,T=1}^{S}(ab,cd))
\right]\;,
\end{eqnarray}
\begin{eqnarray}
E_{K,T=1}(ac,bd) & = & -(-1)^{j_b+j_c} \sum_J (-1)^J (2J+1)
\left\{ \matrix{ j_a & j_b & J \cr j_d & j_c & K \cr}\right\}
\sqrt{(1+\delta_{ab})(1+\delta_{cd})}\nonumber\\
& & \times
{1 \over 4} \left(
V^A_{J,T=0}(ab,cd) -V^S_{J,T=1}(ab,cd)
\right).
\end{eqnarray}

In this isospin formalism, since $A_{JT;MTz}(ab) = (-1)^{j_a+j_b
-J+T} A_{JT;MTz} (ba)$, the definitions of the symmetric and
antisymmetric parts of $V_{JT}^{N} (ab,cd)$, $V^S_{JT} (ab,cd)$, and
$V^A_{JT} (ab,cd)$ become
\begin{equation}
V^{S/A}_{JT} (ab,cd) \equiv
{1 \over 2}
\left[
V^{N}_{JT} (ab,cd) \pm (-1)^{J +j_a+j_b +T-1}V^{N}_{JT}(ba,cd)
\right]\;.
\end{equation}

Note that these expressions allow less freedom in manipulating the
decomposition since we have to couple proton with proton and neutron
with neutron in forming the density operators. Also note that
$E_{K,T=0}(ac,bd)-E_{K,T=1}(ac,bd)$ is an invariant related only to
the physical part of the interactions,
$(V_{J,T=1}^{A}+V_{J,T=0}^{A})$. We can choose all $E_{K,T=1}$ to be
zero in the above (by setting $V_{J,T=1}^{S}= V_{J,T=0}^{A}$) leaving
$E_{K,T=0}$ completely determined by the physical matrix elements. In
that case, we can halve the number of fields to be integrated.
However, while introducing the isovector densities requires more
fields, it also gives more freedom in choosing the unphysical matrix
elements to optimize the calculation.

If we now diagonalize the $E_{KT}(i,j)$ as before and form the
operators
\begin{equation}
\hat{Q}_{KM,T}(\alpha)\equiv {1 \over \sqrt{2(1+\delta_{M,0})}}
(\hat{\rho}_{KM,T}(\alpha)+(-1)^{M}\hat{\rho}_{K-M,T}(\alpha))\;,
\end{equation}
\begin{equation}
\hat{P}_{KM,T}(\alpha)\equiv -{i\over \sqrt{2(1+\delta_{M,0})}}
(\hat{\rho}_{KM,T}(\alpha)-(-1)^{M}\hat{\rho}_{K-M,T}(\alpha))\;,
\end{equation}
the two-body part of the Hamiltonian can finally be written as
\begin{equation}
\hat{H}_2' = {1\over 2}
\sum_{KT} \sum_\alpha
\lambda_{KT}(\alpha) \sum_{M \ge 0} \left(
\hat{Q}^2_{K \,M,T}(\alpha) + \hat{P}^2_{K\, M,T}(\alpha) \right)\;.
\end{equation}

In this decomposition, the one-body hamiltonian $\hat{h}_\sigma$ of
the HS transformation does not mix protons and neutrons. We can then
represent the proton and neutron wavefunctions by separate
determinants, and the number of neutrons and protons will be
conserved rigorously during each Monte Carlo sample. For general
interactions, even if we choose nonzero $E_{K,T=1}$ matrix elements,
the number of fields involved is half that for the neutron-proton
mixing decomposition, and the matrix dimension is also halved. These
two factors combine to speed up the computation significantly. In
this sense, an isospin formalism is more favorable, although at the
cost of limiting the degrees of freedom embodied in the symmetric
matrix elements $V_J^S$.

\subsection{Pairing decomposition}

In nuclei where the pairing interaction is important, it is natural
to cast at least part of the two-body interaction as a quadratic form
in pair creation and annihilation operators. We demonstrate this for
the case where the Hamiltonian is written in the isospin formalism.
Upon diagonalizing $V_{JT}^{A}(ab,cd)$ in Eq.~(3.22), we can write
\begin{equation}
\hat{H}_{2}=\sum_{J T \alpha} \lambda_{J T}(\alpha)\sum_{M T_z} A_{J
T;M T_z}^{\dagger}
(\alpha) A_{J T;M T_z}(\alpha)\;,
\end{equation}
where
\begin{equation}
A^{\dagger}_{J T:M T_z}(\alpha)=\sum_{i} v_{J T\alpha}(i)A^{\dagger}_
{J T, M T_z}(i)\;.
\end{equation}
Separating $A^{\dagger} A$ into commutator and anticommutator terms,
we have
\begin{equation}
\hat{H}_{2}=\hat{H}_{2}'+\hat{H}_{1}'\;;
\end{equation}
\begin{equation}
\hat{H}_{1}'={1 \over 2}\sum_{J T \alpha}\lambda_{J T}(\alpha)
\sum_{M T_z}
\left[ A_{J T;M T_z}^{\dagger}(\alpha),A_{J T;M
T_z}(\alpha)\right]\;;
\end{equation}
\begin{equation}
\hat{H}_{2}'={1\over 2}\sum_{J T \alpha}\lambda_{J T} (\alpha)
\sum_{M T_z} \left\{A_{j T; M T_z}(\alpha)^{\dagger},A_{J T;M
T_z}(\alpha)\right\}\;.
\end{equation}
Clearly, $\hat{H}_{1}'$ is a one-body operator that can be combined
with $\hat{H}_{1}$. The remaining two-body term can be written as a
sum of squares by defining
\begin{mathletters}
\begin{equation}
Q_{J T; M T_z}(\alpha)\equiv {{1}\over{\sqrt{2}}}
\left(A^{\dagger}_{J T;M T_z}(\alpha)+A_{J T; M
T_z}(\alpha)\right)\;;
\end{equation}
\begin{equation}
P_{J T; M T_z}(\alpha)\equiv -{{i}\over{\sqrt{2}}}
\left(A^{\dagger}_{J T;M T_z}(\alpha)-A_{J T;M T_z}(\alpha)
\right)\;,
\end{equation}
\end{mathletters}
so that
\begin{equation}
\hat{H}'_{2}={{1}\over{2}}\sum_{J T} \lambda_{J T}(\alpha)
\sum_{M T_z} \left(Q_{J T; M T_z}^{2}(\alpha)+P_{J T; M
T_z}^{2}(\alpha)
\right)\;.
\end{equation}
As in the density decomposition, we can then couple the $\sigma$ and
$\tau$ fields to $Q$ and $P$, respectively.

Note that in the pairing decomposition, the one-body Hamiltonian
$h(\tau)$ used in the path integral is a generalized one-body
operator that includes density, pair-creation and pair-annihilation
operators. The wavefunction is then propagated as a
Hartree-Fock-Bogoliubov state, rather than as a simple Slater
determinant.

In this decomposition, neutrons and protons are inevitably mixed
together in the one-body Hamiltonian $\hat{h}_\sigma$ (consider the
$Q$,$P$ terms for $T=0$). In fact, $\hat{h}_\sigma$ also does not
conserve the total number of nucleons; rather the conservation is
only statistical after a large number of Monte Carlo samples.

For simplicity, we have described how to decompose the Hamiltonian
solely in density operators or solely in pair operators. However, it
is straightforward to mix the two decompositions with the choice
depending on the type of interactions involved. Consider the `Pairing
plus Quadrupole' model, namely
\begin{equation}
\hat{H}_2= -g P^{\dagger}P -{1 \over 2} \chi Q \cdot Q\;,
\end{equation}
where $P^{\dagger},P$ are the monopole pair creation and annihilation
operator and $Q$ is the quadrupole-moment operator,
\begin{equation}
P^{\dagger}=\sum_{\alpha}
a_{\alpha}^{\dagger}\tilde{a}_{\alpha}^{\dagger}\;,
\end{equation}
\begin{equation}
Q_{\mu}=\sum_{\alpha \beta}\left\langle \alpha | Q_{\mu}
|\beta\right\rangle
a_{\alpha}^{\dagger} a_{\beta}\;.
\end{equation}
Naively, it would be most convenient to use a pairing decomposition
for the pairing interaction and a density decomposition for the
quadrupole interaction. This would require only 8 fields for each
time slice in the HS transformation. The pure pairing or pure density
decompositions would be much more complicated, as the former would
require rewriting the quadrupole interactions in terms of pair
operators and the latter would require rewriting the pairing
interaction in terms of density operators. The numerical examples
given in Section~VI below will illustrate the behavior of such
Hamiltonians under different decompositions.

\section{Calculation of static observables}

In the previous sections, we expressed the evolution operator for a
two-body Hamiltonian as a path integral over auxiliary fields in
which the action involves only density and pair operators. In this
and the following section, we show how to extract observables from
this path integral. The merit of this method will be clear from the
compact formulae involved, which require handling only relatively
small matrices of dimension $N_s$ or $2 N_s$, depending on the type
of decomposition used. We derive formulae for three different
approaches that use the evolution operator to obtain information
about the system: the zero-temperature formalism, the grand-canonical
ensemble, and the canonical ensemble. The zero-temperature and
grand-canonical ensemble methods have been applied, with the density
decomposition, to other physical systems such as the Hubbard Model
\cite{hirsch83}. However, we believe that the canonical ensemble
treatment presented here is novel.

The zero-temperature approach can be used only to extract
ground-state information. On the other hand, the grand-canonical
ensemble allows finite-temperature calculations, but fluctuations in
particle number can be very significant in a system with a small
number of nucleons. Thus, the canonical ensemble is particularly
important in nuclear systems, where the particle number is small and
shell structure is prominent. The grand-canonical ensemble yields
information averaged over neighboring nuclei, which can have very
different properties.

The canonical ensemble is more difficult than the other two
approaches in two respects. First, the canonical (fixed-number) trace
of $U_\sigma$ is more difficult to compute than the wavefunction
overlap of the zero-temperature formalism or the grand-canonical
trace. Second, observables are more difficult to extract since Wick's
theorem does not apply. We suggest two different methods to handle
the canonical ensemble: the activity expansion and the integration
over real coherent states. The coherent-state integration method can
be applied to calculate canonical trace of any particle number
involving only a negligible increase in computation time relative to
the zero-temperature and grand-canonical approaches. Unfortunately,
its utility is limited by the sign problem. On the other hand, the
activity expansion, which is well-suited (and numerically stable) for
calculating nuclei with a small number of valence particles or holes,
is less susceptible to the sign problem. (Yet a third method for
extracting the canonical trace, based on Fourier analysis, shows good
``sign'' statistics and is numerically stable for mid-shell
calculations; it is therefore suitable when the activity expansion
fails. Details of this method will be published elsewhere
\cite{dean93}.)

For each approach, we also derive the general formulae when pair
operators (as well as density operators) are present in the
single-particle Hamiltonian; i.e.,~when a pairing decomposition is
used for some or all of the interaction. As far as we know, there is
no known general formulae for the pairing decomposition formalism.
Using the fermion coherent state formalism \cite{ber66}, we derive in
Appendix~A a set of formulae for the calculation of observables that
are similar to the well-known formulae for a pure density
decomposition. Thus, our methods can be extended to calculations
using general one-body operators in the HS transformation by only
doubling the dimension of the matrices involved.

\subsection{Zero-temperature formalism}

We begin with `zero-temperature' observables. The trial wavefunction
$\psi_0$ in Eq.~(2.4) is, in principle, any state not orthogonal to
the ground state. In practice, it is most conveniently a
Slater-determinant. If we have $N_s$ {\it m}-scheme orbitals
available and $N_v$ indistinguishable particles, $\psi_0$ is
constructed from $N_v$ single-particle wavefunctions, each of which
is a vector with $N_s$ components, and we write ${\psi}_0$ as a
Slater-determinant of a $N_v \times N_s$ matrix, $\bbox{\Psi}_0$.

In the pure density decomposition, consider the one-body Hamiltonian
$\hat{h}=M_{ij}a_i^{\dagger}a_j$ (sums over the indices $i,j$ from 1
to $N_s$ are implicit). The evolved wavefunction,
\begin{equation}
\left| \psi (\Delta \beta) \right\rangle=\exp \left(-\Delta
\beta \hat{h}\right)\left|\psi_0 \right\rangle
\end{equation}
is then a Thouless transformation of the original determinant, the
new state being represented by the matrix $\exp(-\Delta \beta
M)\bbox{\Psi}_0$. We can therefore represent the product of evolution
operators by $N_s \times N_s$ matrices
\begin{equation}
\exp \left(-\Delta \beta \hat{h}_n \right)\ldots\exp \left(-\Delta
\beta \hat{h}_1 \right) \rightarrow
\exp \left(-\Delta \beta M_n \right)\ldots\exp \left(-\Delta \beta
M_1
\right)\;.
\end{equation}

Let us now consider the overlap function in Eq.~(2.16). Let
$U_\sigma(\tau_2,\tau_1)$ be the matrix representing the evolution
operator $\hat{U}_{\sigma}(\tau_2,\tau_1).$ Choosing some value
$\tau$ of the imaginary time at which to insert the operator
$\hat{O}$, we introduce the right and left wavefunctions ${
\psi}_{R,L}(\tau)$ defined by
\begin{mathletters}
\begin{eqnarray}
\left|\psi_R(\tau) \right\rangle & = & \hat{U}_\sigma(\tau,0)\left|
\psi_0 \right\rangle\;,\\
\left|\psi_L(\tau) \right\rangle
& = & \hat{U}^{\dagger}_{\sigma}(\beta,\tau)
\left|\psi_0 \right\rangle\;.
\end{eqnarray}
\end{mathletters}
Then the required overlap in Eq.~(2.16) is
\begin{equation}
\left\langle \psi_0 \left|\hat{U}_\sigma(\beta,0) \right|\psi_0
\right
\rangle = \left\langle \psi_L|\psi_R \right\rangle
=\det \left[\bbox{\Psi}_{L}^{\dagger} \bbox{\Psi}_R \right]\;,
\end{equation}
where
\begin{equation}
\bbox{\Psi}_R(\tau) \equiv \bbox{U}_\sigma(\tau,0)\bbox{\Psi}_0\;;
\quad
\bbox{\Psi}_L(\tau) \equiv \bbox{U}^\dagger_{\sigma}(\beta,
\tau)\bbox{\Psi}_0
\end{equation}
are the matrices representing $\psi_R$ and $\psi_L$. Note that if
there are two distinguishable species of nucleons --- protons and
neutrons --- and we use a decomposition that conserves the numbers of
neutrons and protons, then there is a separate determinant for each
set of single-particle wavefunctions and the total overlap is the
product of the proton and neutron overlaps.

With the basic overlap in hand, we now turn to the expectation value
of an operator $\hat{O}$ for a given field configuration (i.e.,
Eqs.~(2.15) and (2.19)). By Wick's theorem, the expectation value of
any $N$-body operator can be expressed as the sum of products of
expectation values of one-body operators. Hence, the basic quantity
required is
\begin{equation}
\left\langle a_b^{\dagger} a_a\right\rangle_\sigma=
\left[ \bbox{\Psi}_R \left( \bbox{\Psi}^\dagger_L\bbox{\Psi}_R
\right)^{-1}\bbox{\Psi}^\dagger_L\right]_{ba}\;.
\end{equation}
(A straightforward derivation of this expansion can be found in
\cite{loh92}.) Again, when the decomposition separately conserves
proton and neutron numbers, the expectation values for proton and
neutron operators are given by separate matrices.

These formulae can be extended to the case where pair operators are
present in the one-body Hamiltonian; i.e., when $\hat{h}$ is of the
form
\begin{equation}
\hat{h}=\sum_{i j=1}^{N_s} (\Theta_{ij}a_{i}^{\dagger}a_{j}
+\Delta_{ij}a_{i}^{\dagger}a_{j}^{\dagger}+\Lambda_{ij}a_{i}a_{j})\;,
\end{equation}
where $\Theta, \Delta$, and $\Lambda $ are general $N_s \times N_s$
complex matrices. Our first task is to find a simple expression for
$\langle\psi_{0}|\hat{U}_\sigma(\beta,0)|\psi_{0}\rangle$ where
$\hat{U}_\sigma(\beta,0)= \prod_{n=1}^{N_t} \exp(-\Delta \beta
\hat{h}_{\sigma_n})$. Using the fermion coherent state representation
and Grassman algebra, we derive the expressions in Appendix~A. Here
we simply state the results.

If the trial wavefunction $\psi_0$ is a quasi-particle vacuum, such
that $\beta_{i} |\psi_{0}\rangle =0$ where $\beta_{i}= \sum_j
u_{ij}a_{j} +v_{ij}a_{j}^{\dagger}$, then
\begin{equation}
\langle \psi_0|\prod_{n=1}^{N_t} \exp  (-\Delta \beta
\hat{h}_{n})|\psi_{0}\rangle
 = \det\left[\pmatrix{v^{*} & u^{*}\cr}
\bbox{U}_\sigma(\beta,0)\pmatrix{v^{T}\cr
u^{T}\cr}\right]^{1\over 2}
\exp \left(-{{\Delta \beta} \over 2} \sum_{n=1}^{N_{t}} {\rm Tr}
[\Theta_{n}]
\right)\;,
\end{equation}
where
\begin{equation}
\bbox{U}_\sigma(\beta,0)=\exp(-\Delta \beta
M_{N_t})\ldots\exp(-\Delta
\beta M_1)\;,
\end{equation}
is the matrix representing the evolution operator
$\hat{U}_\sigma(\beta,0)$, and $M_n$ is the $2N_s \times 2N_s$ matrix
representing $\hat{h}_{n}$:
\begin{equation}
M_n=\pmatrix{\Theta_n & \Delta_n-\Delta_{n}^{T}\cr
\Lambda_n-\Lambda_{n}^{T} & -\Theta_{n}^{T}\cr}\;.
\end{equation}

Here, the evolution operator $\hat{U}_\sigma(\tau_{2},\tau_{1})$ is
represented by a $2N_s \times 2N_s$ matrix and the many-body
wavefunction is represented by a $2N_s \times N_s$ matrix,
independent of the number of particles present. In analogy to
Eqs.~(4.3--4.6), we can write
\begin{equation}
\left\langle \psi_{0}\left|\hat{U}_\sigma(\beta,0) \right
|\psi_{0}\right\rangle =
\left\langle \psi_{L}| \psi_{R}\right\rangle
= \det\left[\bbox{\Psi}_{L}^{\dagger} \bbox{\Psi}_{R}
\right]^{1\over2}
\exp \left(-{{\Delta \beta}
\over 2}
\sum_{n=1}^{N_t} {\rm Tr}[\Theta_{n}]\right)
\end{equation}
where
\begin{equation}
\bbox{\Psi}_R \equiv \bbox{U}_\sigma(\tau,0)\pmatrix{v^T\cr
u^T\cr},\bbox{\Psi}_L
\equiv \bbox{U}_\sigma^{\dagger}(\beta,\tau) \pmatrix{v^T\cr
u^T\cr}\;.
\end{equation}

To calculate the expectation value of a generalized one-body
operator, we proceed as in the pure density case. Let
\begin{mathletters}
\begin{equation}
\alpha_i= a_i,\quad i=1,\ldots,N_s
\end{equation}
\begin{equation}
\alpha_{i+N_s}= a_{i}^{\dagger},\quad i=1,\ldots,N_s\;.
\end{equation}
\end{mathletters}
Then the one-quasiparticle density matrix is
\begin{eqnarray}
\left\langle \alpha_{a}^{\dagger} \alpha_{b} \right\rangle_\sigma
& = & { {\left\langle \psi_{L}\left|\hat{\rho}_{ab}\right
|\psi_{R}\right
\rangle} \over {\left\langle \psi_{L}|\psi_{R}\right
\rangle}}\nonumber\\
& = & {1 \over 2} \left[ \bbox{\Psi}_R(\bbox{\Psi}_L^{\dagger}
\bbox{\Psi}_R
)^{-1} \bbox{\Psi}_{L}^{\dagger}\right]_{ba}
-{1 \over 2}\left[\pmatrix{\bbox{0} & \bbox{1}\cr \bbox{1} &
\bbox{0}\cr}
\bbox{\Psi}_R
(\bbox{\Psi}_L^\dagger \bbox{\Psi}_R)^{-1}\bbox{\Psi}_L^\dagger
\pmatrix{\bbox{0} & \bbox{1} \cr \bbox{1} & \bbox{0}\cr}
\right]_{ab} +\delta_{ab}\nonumber\\
& = & \left[\bbox{\Psi}_{R}(\bbox{\Psi}_{L}^{\dagger}
\bbox{\Psi}_{R})^{-1}\bbox{\Psi}_{L}^{\dagger}\right]_{ba}.
\end{eqnarray}
The final step follows from the properties of $\bbox{U}$ in
Eq.~(A20). Note that both the overlap and the Green's function are
similar to those of the density decomposition. However, the enlarged
dimension of the representation matrices causes the overlap to be the
square-root of a determinant rather than just a determinant. We know
of no simple way to determine the sign of the square root.
Computationally, it can be traced by watching the evolution of
$\langle \psi_{0}|\hat{U}_\sigma(\tau,0) |\psi_0\rangle $ as
$\hat{U}_\sigma$ is built up from $0$ to $\beta$ (Appendix~B),
although at the expense of increased computation. Also, the linear
dimension of the matrix used in the calculation is twice that for the
pure density decomposition case so that the pairing decomposition is
more computationally demanding. However, when the interaction has a
strong pairing character, it has the potential for a more effective
Monte Carlo sampling, and offers greater freedom in the
decomposition, which might be used to mitigate the sign problem.

\subsection{Grand-Canonical Ensemble}

For the grand-canonical ensemble, the trace in Eq.~(2.18) is a sum
over all possible many-body states with all possible nucleon numbers,
and a chemical potential in $\hat{H}$ is required. For the pure
density decomposition, the many-body trace is given by
\begin{equation}
\zeta(\sigma)=
{\rm \hat{T}r\,} \hat{U}_{\sigma}(\beta,0)
= \det \left( \bbox{1} + \bbox{U}_\sigma(\beta,0) \right)\;,
\end{equation}
which can be proved by expanding the determinant \cite{loh92}.

For the expectation value of a one-body operator, one can
recapitulate exactly the argument of the zero-temperature development
and obtain
\begin{equation}
\left\langle a_a^{\dagger} a_b \right\rangle_\sigma
= \left[ \left( \bbox{1} + \bbox{U}_\sigma(\beta,0) \right)^{-1}
\bbox{U}_\sigma(\beta,0) \right]_{ba}\;.
\end{equation}

We have extended these formulae to decompositions that involve
pairing operators (Appendix~A). The results are given in terms of the
$2N_s \times 2N_s$ matrices $\bbox{M}_n,\bbox{U}(\beta,0)$
representing the Hamiltonians $\hat{h}_n$ and the evolution operator
$\hat{U}(\beta,0)$, (Eqs.~(4.9,10)) namely
\begin{equation}
{\rm \hat{T}r}\left[\hat{U}(\beta,0) \right]
=\det \left[1+\bbox{U}(\beta,0) \right]^{1\over 2}
\exp \left(-{{\Delta \beta}\over 2}{\rm Tr}[\Theta_n] \right)\;.
\end{equation}

To motivate this formula, consider the simplest case of one time
slice where $\hat{U}=\exp(-\hat{h})$ and $\hat{h}$ is hermitian and
in the form (4.7). Then, $\hat{h}$ can be diagonalized by a HFB
transformation
\begin{mathletters}
\begin{eqnarray}
\hat{h}&=&{1 \over 2} \pmatrix{a^{\dagger} & a\cr}
\pmatrix{\Theta & \Delta-\Delta^{T} \cr
\Lambda-\Lambda^{T} & -\Theta^{T} \cr}
\pmatrix{a & a^{\dagger}\cr} +{1 \over 2} {\rm Tr}[\Theta]\\
&=&\sum_{i}e_{i}\beta_{i}^{\dagger} \beta_{i}- {1\over 2}e_i
+{1\over 2} {\rm Tr}[\Theta]
\end{eqnarray}
\end{mathletters}
where
\begin{equation}
\pmatrix{\beta\cr \beta^{\dagger}\cr}=
\pmatrix{u & v^{*}\cr
v & u^{*} \cr} \pmatrix{a\cr a^{\dagger}\cr}
\end{equation}
and
\begin{eqnarray}
\pmatrix{u & v^{*}\cr
v & u^{*} \cr} & &
\pmatrix{\Theta & \Delta-\Delta^{T} \cr
\Lambda-\Lambda^{T} & -\Theta^{T} \cr}
\pmatrix{u^{\dagger} & v^{\dagger}\cr v^{T} & u^{T}\cr}\nonumber\\
& & =\pmatrix{\epsilon_{1} & 0 & \ldots & 0 & 0 & \ldots\cr
0 & \epsilon_{2} & \ldots & 0 & 0 & \ldots\cr
\vdots & \vdots & \ddots & \vdots & \vdots & \ddots\cr
0 & 0 & \ldots & -\epsilon_{1} & 0 & \ldots\cr
0 & 0 & \ldots & 0 & -\epsilon_{2} & \ldots\cr
\vdots & \vdots & \ddots & \vdots & \vdots & \ddots\cr}\;.
\end{eqnarray}
In the diagonal form, ${\rm \hat{T}r}[\exp(-\hat{h})]$ can be
identified easily as
\begin{equation}
\prod_{i}[(1+e^{-\epsilon_{i}})e^{{1\over 2}\epsilon_i}]
e^{-{1\over 2}{\rm Tr}[\Theta]}
=\prod_i\sqrt{(1+e^{-\epsilon_{i}})(1+e^{ \epsilon_i})}
e^{-{1\over 2}{\rm Tr}[\Theta]}
=\det[(1+e^{-M})^{1 \over 2}] e^{-{1\over 2} {\rm Tr}[\Theta]}\;,
\end{equation}
due to the invariance of the determinant with respect to similarity
transformations. Here, as in the overlap formulae for
zero-temperature approach, the grand-canonical trace is given as the
square root of a determinant, so that the evolution of the sign is
important (Appendix~B). The formula for the observables can be
calculated from
\begin{eqnarray}
\left\langle \alpha_{a}^{\dagger}\alpha_b \right\rangle_\sigma
& = & {1\over 2}
\left[\left(1+\bbox{U}(\beta,0)\right)^{-1}\bbox{U}(\beta,0)\right]_{
ba}\nonumber\\
& &
-{1\over 2}\left[
\pmatrix{\bbox{0} & \bbox{1} \cr \bbox{1} & \bbox{0} \cr}
(bbox{1}+ \bbox{U}
(\beta,0))^{-1}\bbox{U}(\beta,0)
\pmatrix{\bbox{0} & \bbox{1}\cr \bbox{1} & \bbox{0} \cr}
\right]_{ab}+{1 \over 2}\delta_{ab}\nonumber\\
& = &
\left[\left(1+\bbox{U}(\beta,0)\right)^{-1}\bbox{U}(\beta,0)\right
]_{ba}\;.
\end{eqnarray}

\subsection{Activity expansion for the canonical ensemble}

As mentioned in the beginning of this section, the grand-canonical
ensemble may lead to large fluctuations in the particle number for
systems with few particles, and so is particularly ill-suited for
small nuclear systems, and although the particle number does not
fluctuate in the zero-temperature approach, that formalism can only
give ground-state results. The canonical ensemble is therefore
important for studying thermal behavior, as well as the ground state
of large systems.

In the canonical ensemble, we have to find the trace of
$\hat{U}_\sigma(\beta,0)$ over all states with a fixed particle
number $N_v$ (actually fixed proton and neutron numbers). We discuss
two methods of achieving this: the activity expansion presented here
and the integration over real coherent states presented in the
following subsection.

Consider first the case when only density operators are present in
the one-body Hamiltonian $\hat{h}$. From the grand-canonical ensemble
formulae (4.15), one can see that the canonical trace is just the sum
of all the $N_v\times N_v$ sub-determinants. More explicitly, we
consider the {\it activity expansion}: for some parameter $\lambda$,
let
\begin{equation}
Z(\beta,\lambda) = {\rm \hat{T}r} \lambda^{\hat{N}_v}
e^{-\beta\hat{H}}
= \sum_{N_v} \lambda^{N_v} \, Z_{N_v}(\beta)\;.
\end{equation}
In our matrix representation,
\begin{equation}
{\rm \hat{T}r}\, \lambda^{\hat{N}} \hat{U}_\sigma(\beta,0)
= \det ( \bbox{1} + \lambda \bbox{U})\;.
\end{equation}
If Eq.~(4.24) is expanded in powers of $\lambda$, the coefficient of
$\lambda^{N_v}$ is just the canonical trace of
$\hat{U}_\sigma(\beta,0)$ over $N_v$ particles, so that
$\det(\bbox{1}+\lambda \bbox{U})$ is the generating function for the
canonical trace. Thus,
\begin{equation}
Z_{N_v}(\beta)=\int{\cal D}[\sigma]G(\sigma)\zeta_{N_v}(\sigma)\;,
\end{equation}
where
\begin{equation}
\det\left(1+\lambda \bbox{U}_\sigma(\beta,0)\right)=\sum_{N_v}
\lambda^{N_v}\zeta_{N_v}(\sigma)\;.
\end{equation}

The trick now is to find simpler expressions for $\zeta_{N_v}$,
instead of doing the explicit sum over the determinants. To do this,
write
\begin{equation}
\det (\bbox{1} + \lambda \bbox{U})
= \exp \, {\rm Tr \,} \ln
(\bbox{1} + \lambda \bbox{U})
= \exp \left(\sum_{n=1}^\infty{ {(-1)^{n-1}}\over n} \lambda^n {\,
\rm Tr \,}[
\bbox{U}^n]\right)\;.
\end{equation}
This expansion converges to the generating function because
$Z(\beta,\lambda)$ is a polynomial in $\lambda$ of finite order
(i.e., $N_v$ can be at most $N_s$). The coefficient of $\lambda^n$ in
the exponential is readily found. For a given particle number $N_v$,
we need only calculate matrix traces up to ${\rm Tr}[\bbox{U}^{N_v}]$
and the coefficient of $\lambda^{N_v}$ can be extracted accordingly.

For one-body expectation values, using again the grand-canonical
trace as the generating function for calculating observables and
collecting all terms with coefficient $\lambda^{N_v}$, we arrive at
\begin{equation}
\left\langle a_a^{\dagger}a_b \right\rangle_{\sigma,N_v} = \sum_{n
=1}^{N_v} (-1)^{n-1} \left( \bbox{U}^n \right)_{ba}
\, \zeta_{N-n}(\sigma) / \zeta_N(\sigma)\;.
\end{equation}
The expectation value of two-body operators, $\left\langle
\rho_{ab}\rho_{cd} (\sigma) \right\rangle_{N_v} $, is nontrivial as
Wick's theorem must be modified, but again one can simply collect the
terms with coefficient $\lambda^{N_v}$ and obtain
\begin{eqnarray}
\left\langle a_a^{\dagger}a_b a_c^{\dagger} a_d
\right\rangle_{\sigma,N_v}
& = & \sum_{n=1}^{N_v}
\left\{ \sum_{m=1}^{N_v} \left[ (-1)^{m+n}
\left( \bbox{U}^n_{ba} \bbox{U}^m_{dc} -\bbox{U}^n_{da}
\bbox{U}^m_{bc}
\right)
\zeta_{N_v-m-n}(\sigma) / \zeta_{N_v}(\sigma) \right
]\right.\nonumber\\
& & \qquad + \delta_{bc}(-1)^{N_v-1} \bbox{U}^n_{da}
\zeta_{N_v-n}(\sigma) / \zeta_{N_v}(\sigma) \Biggl \}\;.
\end{eqnarray}

This form of the activity expansion works best for $N_v \le N_s/2$
(mostly empty model spaces). When $N_v > N_s/2$ (mostly filled
spaces), it is more efficient to expand in the activity of the hole
states. In this case, we define
\begin{equation}
Z_{\sigma}(\beta,\lambda)={\rm \hat{T}r}[\lambda^{N_s-\hat{N}}
e^{-\beta \hat{H}}]\;,
\end{equation}
and as in Eq.~(4.24),
\begin{equation}
{\rm \hat{T}r}[\lambda^{N_s-\hat{N}}\hat{U}(\beta,0;\sigma)]
=\det[\lambda +\bbox{U}]\;.
\end{equation}
The coefficient of $\lambda^{N}$ is the partition function for $N$
holes,
\begin{equation}
\det[\lambda +\bbox{U}]=
\det[\bbox{U}] \exp(\sum_{n\geq 1} {1 \over n} (-1)^{n-1} \lambda^{n}
Tr[\bbox{U}^{-n}])\equiv
\sum_{N} \lambda^{N} \zeta_{N}'\;,
\end{equation}
the expectation value of a one-body operator is
\begin{equation}
\langle\rho_{ij}\rangle_{N \ {\rm holes}} =\sum_{n=0}^{N} (-1)^{n}
\bbox{U}_{ji}^{-n}
\zeta_{N-n}'/\zeta_{N}'\;,
\end{equation}
and the expectation of a two-body operator is
\begin{eqnarray}
\langle\rho_{ij} \rho_{kl}\rangle _{N \ {\rm holes}} & = &
\sum_{n=0}^{N}
\delta_{jk}(-1)^{n} \bbox{U}_{li}^{-n} \zeta_{N-n}'/\zeta_{N}')
\nonumber \\
& & +(\sum_{m=0}^{N}
(-1)^{m+n} (\bbox{U}_{ji}^{-n} \bbox{U}_{lk}^{-m} -\bbox{U}_{li}^{-n}
\bbox{U}_{jk}^{-m})\zeta_{N-m-n}'
/\zeta_{N}' \;.
\end{eqnarray}

When pairing operators are involved, we use Eq.~(4.17) for the
grand-canonical trace, which becomes the generating function for the
corresponding canonical trace. As an illustration, we display the
formula for the expansion in particle activity,
\begin{eqnarray}
{\rm \hat{T}r}\left[\lambda^{\hat{N}}\hat{U}\right] & = & \det
\left[\pmatrix{ \bbox{1} & \bbox{0} \cr \bbox{0} & \bbox{1}\cr}+
\pmatrix{\lambda \bbox{1} & \bbox{0}\cr \bbox{0} & {1\over
\lambda}\bbox{1}\cr}
\bbox{U} \right]^{1\over 2} \lambda^{N_s \over 2} \nonumber\\
& = & \det\left[\pmatrix{\bbox{1} & \bbox{0}\cr \bbox{0} & \bbox{1}
\cr}+
\pmatrix{\lambda \bbox{1} & \bbox{0}\cr\bbox{0} & {1\over
\lambda}\bbox{1}\cr}
\pmatrix{\bbox{S^{11}} & \bbox{S^{12}} \cr \bbox{S^{21}} &
\bbox{S^{22}}\cr} \right]^{1 \over 2} \lambda^{N_s \over 2}\nonumber
\\
& = & \det\left[\pmatrix{\bbox{1} & \bbox{S^{12}}\cr \bbox{0} &
\bbox{S^{22}}\cr}\right]
^{1\over 2} \det \left[\pmatrix{\bbox{1} & \bbox{0}\cr \bbox{0} &
\bbox{1}\cr} +
\lambda \pmatrix{\bbox{1} & \bbox{S^{12}}\cr \bbox{0} &
\bbox{S^{22}}\cr
}^{-1}\pmatrix{\bbox{S^{11}} & \bbox{0}\cr
\bbox{S^{21}} & \bbox{1}\cr}\right]^{1 \over 2}\nonumber\\
& = & \det \left[\pmatrix{\bbox{1} & \bbox{S^{12}}\cr
\bbox{0} & \bbox{S^{22}}\cr}\right]^{1\over 2} \exp
\left({1\over 2}{\rm Tr} \left[\ln \left(
\pmatrix{\bbox{1} & \bbox{0}\cr \bbox{0} & \bbox{1}\cr}+\lambda
\bbox{Y}\right)
\right]\right)\nonumber\\
& = & \det \left[\bbox{S^{22}}\right]^{1\over 2}\exp\left({1\over
2}\sum_{n} {\rm Tr}\left[\bbox{Y}^{n}\right](-1)^{n-1}{\lambda^{n}
\over n}\right),
\end{eqnarray}
where the definitions of $\bbox{S^{11}, S^{12}, S^{21}, S^{22}}$ are
obvious, and
\begin{equation}
\bbox{Y} \equiv \pmatrix{\bbox{1} & \bbox{S^{12}} \cr \bbox{0} &
\bbox{S^{22}}
\cr}^{-1}
\pmatrix{\bbox{S^{11}} & \bbox{0} \cr \bbox{S^{21}} & \bbox{1}\cr}\;.
\end{equation}
The expectation values of one- and two-body operators can then be
derived as in the pure density decomposition.

\subsection{Canonical ensemble via coherent states}

We make use of the operator
\begin{equation}
\hat{J}=C \int \prod_{ph} d X_{ph}{{ |X\rangle \langle X|}\over
{\det(1+X^{T}X)^{{N_s \over 2}+1}}}\;,
\end{equation}
which can be shown \cite{dob83} to be a resolution of unity in the
Hilbert space of $N_v$ fermions occupying $N_s$ levels. Here,
$X_{ph}$ are the real integration variables, $|X\rangle$ are the real
coherent states
\begin{equation}
| X\rangle =\exp(\sum_{hp} X_{ph} a_{p}^{\dagger} a_h) |0\rangle,
\quad h=1,\ldots,N_v;\ p=N_v+1,\ldots,N_s\;,
\end{equation}
$|0\rangle$ is the $N_v$-fermion state with the levels $1,\ldots,N_v$
occupied and $C$ is a normalization constant,
\begin{equation}
C=\prod_{p=N_v+1}^{N_s}C_p,\quad
C_p=\pi^{-N_s/2}\Gamma\left({p\over2}+1\right)/
\Gamma \left({{p-N_v}\over 2}+1\right)\;.
\end{equation}
The canonical trace can then be cast in a form of expectation with
integration over the variables $X_{ph}$. For any operator $\hat{O}$,
\begin{equation}
{\rm \hat{T}r}[\hat{O}] = C \int d X {{\langle X| \hat{O} |X\rangle}
\over
{\det(1+X^T X)^{{N_s \over 2} +1}}}
\end{equation}
where $d X=\prod_{ph}dX_{ph}$. The thermal canonical expectation is
then
\begin{eqnarray}
\langle \hat{O} \rangle & = & {{ {\rm \hat{T}r}[\hat{O} e^{-\beta
\hat{H}}]} \over
{{\rm \hat{T}r}[e^{-\beta \hat{H}}]}}\nonumber\\
& = & {{\int {\cal D}[\sigma]G(\sigma) dX \langle X|
\hat{U}_\sigma(\beta,0)\hat{O} |X\rangle
/\det(1+X^T X)^{{N_s /2} +1}}\over
{\int {\cal D}[\sigma]G(\sigma) d X
\langle X| \hat{U}_\sigma(\beta,0) |X\rangle/\det(1+X^T X)^{{N_s / 2}
+1}}}\;.
\end{eqnarray}

We can compute the overlap and the observable in (4.41) as in the
zero-temperature formalism, but we now have to do the Monte Carlo
integration over the fields $X_{ph}$ as well. However, the number of
these fields scales only as $N_s^{2}$, which is about the same as the
number of $\sigma$ fields in one time slice, so that introducing the
coherent-state integration is not much more computationally expensive
than the zero-temperature formalism. The advantage of coherent states
is that we do not have to find ${\rm Tr}[\bbox{U}(\beta,0)^{n}]$.
However, the integration of the coherent states may aggravate the
sign problem, as will be discussed in Sections~VI and VII.

\section{Dynamical correlations}

In the previous section, we discussed how to extract the expectation
value of a one-body operator $\langle \hat{O} \rangle$ and, by use of
Wick's theorem and its extension to the canonical case, equal-time
two-body operators $\langle \hat{A} \hat{B} \rangle$, etc. A great
deal of the interesting physics, however, is contained in the
dynamical response function $\langle
\hat{O}^\dagger(t)\hat{O}(0)\rangle $ where $\hat{O}(t)=e^{-i \hat{H}
t} \hat{O} e^{i \hat{H}t}$. In our calculations, we evaluate the
imaginary-time response function $\langle \hat{O}^\dagger (i \tau)
\hat{O}(0) \rangle$ and from it deduce the associated strength
function, $f_{\hat{O}}(E)$.

In the zero temperature formalism, the strength function is
\begin{equation}
f_{\hat{O}}(E) \equiv \sum_{f} \delta (E-E_f+E_i) | \langle
f|\hat{O}|i\rangle|^{2}\;,
\end{equation}
where $i$ is the ground state, while in the canonical or
grand-canonical ensemble,
\begin{equation}
f_{\hat{O}}(E) \equiv {1\over Z}\sum_{i,f} \delta(E-E_f
+E_i)e^{-\beta E_i}
|\langle f|\hat{O} |i\rangle|^{2}\;.
\end{equation}
Thus, the imaginary-time response function is related to the strength
function by a Laplace transform,
\begin{equation}
R(\tau)\equiv \langle \hat{O}^{\dagger}(i \tau) \hat{O}(0)\rangle =
\int_{-\infty}^{ \infty} f_{\hat{O}}(E) e^{-\tau E} dE\;.
\end{equation}

Recovering the strength function from $R$ by inversion of the Laplace
transform is an ill-posed numerical problem. Different methods have
been proposed to surmount this difficulty \cite{white89,sil90}. We
resort to making the best ``guess'' for the strength function via the
Classic Maximum Entropy (MaxEnt) method \cite{skill89,gull89}, which
was first introduced to recover strength functions in Monte Carlo
calculations by Silver {\it et al\/.} \cite{sil90} and has since been
widely used in similar condensed matter simulations. It is in essence
a least-squares fit biased by a measure of the phase space of the
strength function.

In MaxEnt methods, the function to be minimized is ${1 \over 2}
\chi^2 -\alpha S$, where $\chi^{2}$ is the usual square of the
residuals, $S$ is the entropy of the phase space (not to be confused
with the action in the auxiliary field path integral), and $\alpha $
is a Lagrange multiplier. In Classic MaxEnt $\alpha$ is determined
self-consistently. The method, described briefly in Appendix~C, also
can yield error estimates for the extracted strength function.

To calculate the imaginary time response, consider $\left\langle
\hat{A}(i \tau) \hat{B} (0)\right\rangle$ and write the thermal
expectation as a path integral:
\begin{eqnarray}
\langle \hat{A}(i\tau) \hat{B}(0) \rangle & = & {{{\rm \hat{T}r}
[e^{-(\beta -\tau)\hat{H}}
\hat{A} e^{-\tau \hat{H}} \hat{B}]} \over {{\rm \hat{T}r}[e^{-\beta
\hat{H}}]}}\nonumber\\
& = & {{\int {\cal D}[\sigma] G(\sigma) {\rm \hat{T}r}
[\hat{U}_\sigma(\beta,0)]
{{{\rm \hat{T}r}[\hat{U}_\sigma(\beta,\tau) \hat{A}
\hat{U}_\sigma(\tau,0)
\hat{B}]}\over {{\rm \hat{T}r}[\hat{U}_\sigma(\beta,0)]}}}\over
{\int {\cal D} [\sigma] G(\sigma) {\rm \hat{T}r}[\hat{U}_\sigma
(\beta,0)]}}\;.
\end{eqnarray}
Upon defining
\begin{equation}
\hat{O}_{\sigma}(\tau)\equiv \hat{U}_\sigma(\tau,0)^{-1} \hat{O}
\hat{U}_\sigma(\tau,0)
\end{equation}
we have an expression suitable for Monte Carlo evaluation,
\begin{mathletters}
\begin{eqnarray}
\langle\hat{A}(i \tau)\hat{B}(0)\rangle
& = & {{\int {\cal D} [\sigma] G(\sigma)
{\rm \hat{T}r}[\hat{U}_\sigma(\beta,0)]
{{{\rm \hat{T}r}[\hat{U}_\sigma(\beta,0)
\hat{A}_\sigma(\tau) \hat{B}_\sigma(0)]} \over {{\rm \hat{T}r}
[\hat{U}_\sigma(\beta,0)]}}} \over {\int{\cal D}[\sigma] G(\sigma)
{\rm \hat{T}r}
[\hat{U}_\sigma(\beta,0)]}}\\[.4cm]
& = & {{\int {\cal D} [\sigma] e^{-{\cal S}(\sigma)} \langle
\hat{A}_\sigma(\tau)
\hat{B}_\sigma(0)\rangle} \over {\int {\cal D} [\sigma ]
e^{-{\cal S}(\sigma)}}}\;.
\end{eqnarray}
\end{mathletters}

We now proceed to find $\hat{O}_\sigma(\tau)$. For the purpose of
illustration, we show the formulae for the pure density decomposition
(formulae for the general case can be derived similarly). For the
simplest case when $\hat{O}=a_{i}^{\dagger}$ or $\hat{O}=a_i$, it can
be shown that \cite{loh92}
\begin{mathletters}
\begin{eqnarray}
{a_{i}^{\dagger}}_\sigma(n \Delta \beta)
& = & \sum_j [\bbox{U}_\sigma(n \Delta \beta,0)^{-1}]^{T}_{ij}
a_{j}^{\dagger}\\
{a_{i}}_{\sigma}(n \Delta \beta) & = &
\sum_j \bbox{U}_\sigma(n\Delta \beta,0)_{ij} a_j\;.
\end{eqnarray}
\end{mathletters}
In this way, the creation and annihilation operators can be
`propagated' back to $\tau=0$. Any operator $\hat{O}_\sigma(\tau)$
can be first expressed in ${a}_\sigma(\tau)$ and
${a^{\dagger}}_\sigma(\tau)$ and therefore can be propagated back and
expressed in $a$ and $a^{\dagger}$. For example, suppose $\hat{O}
=C_{ij}a_{i}^{\dagger}a_j $ is a one-body operator. Then
\begin{eqnarray}
\hat{O}_\sigma(\tau) & = & \hat{U}_\sigma(\tau,0)^{-1} \hat{O}
\hat{U}_\sigma(\tau,0)\\
& = & [\bbox{U}_\sigma(\tau,0)^{-1} \bbox{C}
\bbox{U}_\sigma(\tau,0)]_{ij}a_{i}^{\dagger} a_{j}\;.
\end{eqnarray}
Thus, the response function can be measured in the same way as the
static observables.

\section{Computational details and illustrations}

\subsection{Monte Carlo Methods}

Monte Carlo evaluation of the path integral requires a weight
function. We have tried two different choices for the weight
function, each with advantages and disadvantages.

One choice for the weight function is a Gaussian, with the the static
mean-field solution as the centroid, and the widths given by the RPA
frequencies. Thus, much of the known physics is embodied in the
weight and the Monte Carlo evaluates corrections to the mean-field +
RPA approximation. Further, it is simple to efficiently generate
random field configurations with a Gaussian distribution.

Unfortunately, Gaussian sampling has several disadvantages. First,
finding the RPA frequencies can be extremely time consuming since we
have to calculate and diagonalize the curvature matrix ${{\partial^2}
\over {\partial \sigma_{\alpha,i} \partial \sigma_{\gamma,j}}}
{\partial^2{\cal S}\over \partial\sigma\partial\sigma}$ at the
mean-field solution. (Here, $\alpha$, $\gamma$ run through the number
of quadratic terms in the Hamiltonian and $i$,$j$ run through the
number of time slices.) For any general interaction, the curvature
matrix has a large dimension, $N^2_sN_t$. Second, the Gaussian has to
be modified when there is spontaneous symmetry breaking in the mean
fields (such as quadrupole deformation). Otherwise, the Goldstone
modes in the the RPA frequencies (e.g., zero eigenvalues
corresponding to shape rotations) will destroy the normalizability of
the weight function. Finally, multiple mean-field solutions well
separated from each other can also pose a problem, so that multiple
Gaussians may be needed.

A more satisfactory route is to choose $|\exp(-{\cal
S})|=G(\sigma)|\zeta(\hat{U}_\sigma(\beta,0))|$ as the weight
function and to use a stochastic random walk such as the Metropolis
algorithm to generate the fields. The expectation of an observable is
then given by Eq.~(2.21) where
\begin{equation}
\Phi_i=\exp[-i{\rm Im}({\cal S}_i)]\;.
\end{equation}
The Metropolis algorithm is free of the disadvantages for the
Gaussian weight function, in that it will eventually sample the
entire space where $\vert\exp(-{\cal S})\vert$ is significant. The
main disadvantages of Metropolis are its inefficiency as currently
implemented and the ``sign problem.'' Let us define the denominator
of Eq.~(2.21) by $\langle \Phi\rangle$:
\begin{equation}
\left\langle \Phi \right\rangle \equiv {1 \over N} \sum_i
\exp[-i{{\rm Im}({\cal S}_i)}]\;.
\end{equation}
If $\langle \Phi \rangle \ll 1$, the large fluctuations defeat any
attempt at a Monte Carlo evaluation. This `sign problem' will be
addressed in detail in the examples illustrated below.

For the Metropolis algorithm, we take random steps in the fields
time-slice by time-slice, following a sweeping procedure introduced
by Sugiyama and Koonin \cite{koon82}. For the Monte Carlo results to
be valid, one requires that the points in the random walk be both
distributed as the weight function and be statistically independent.
The first requirement translates into starting the fields in a region
of statistically significant weight; if the initial configuration has
a small weight, a number of initial ``thermalization'' sweeps are
usually needed to relax the fields to this region. The second
requirement means that the walker must sweep through the fields many
times to de-correlate the samples.

The number of thermalization sweeps and decorrelation sweeps
increases with system size, and the choice of random walk procedure
greatly affects the sampling efficiency. In the early stage of
investigation, we allowed all fields $\sigma_\alpha$ at one
time-slice to change with equal probability within a certain step
size. The acceptance is then determined by the ratio of the weight
$\vert\exp(-{\cal S})\vert$ of the old and new configurations, and we
adjusted the step size to give an average acceptance probability of
0.5. The calculations of {\it sd}-shell nuclei described below then
needed approximately 2000 thermalization sweeps and up to 200
decorrelation sweeps between samples. In our later work, we used
another random walk algorithm, in which a fixed number of fields at
one time slice are randomly chosen for updating; those chosen are
generated from the Gaussian distribution in Eq.~(2.9b) while the
others are kept at their previous values, and the acceptance is
determined by the ratio of $\zeta$ in the new and old configurations.
This alternative algorithm needs only some 200 thermalization sweeps
and 10 decorrelation sweeps; i.e., it is 10 times more efficient than
the previous method. In addition, boundaries where $|\exp(-{\cal
S})|=0$ can confine the first random walk algorithm, while they do
not affect the second.

The random walk can thermalize faster if it starts from a
configuration where the weight $W(\sigma)$ is large. Usually we start
from the static mean fields, given by
\begin{mathletters}
\begin{eqnarray}
\sigma_{\alpha,n} & = & \bar{\sigma}_{\alpha},
\quad n=1,N_t\;;\\
\bar{\sigma}_{\alpha} & = & -s_{\alpha}{\rm sign}
(V_{\alpha})\langle\hat{O}_\alpha \rangle_{\bar{\sigma}}\;.
\end{eqnarray}
\end{mathletters}

One can show that for canonical and grand-canonical calculations, the
static mean field $\bar{\sigma}_{\alpha}$ is a saddle point of the
weight function,
\begin{equation}
{{\partial [G({\sigma})\zeta(\sigma)]}
\over {\partial \sigma_{\alpha,i}}}| _{\bar{\sigma}}= 0\;.
\end{equation}
For the zero-temperature approach, we also choose the self-consistent
field solution $\bar{\sigma}_{\alpha}$, although Eq.~(6.4) is not
rigorously satisfied. This is preferable to starting the fields at
zero, which may be far from configurations of significant weight.

The mean field solution (6.3) can be found iteratively
\begin{equation}
\{\bar{\sigma}_{\alpha}\}_{i+1} = -s_{\alpha}{\rm sign}
(V_\alpha)\langle\hat{\cal{O}_\alpha}\rangle_
{\bar{\sigma}_i}\;.
\end{equation}
For the systems we have tested, Eq.~(6.5) usually converges within
100 interactions starting from $\bar{\sigma}_\sigma=0$. For larger
systems and at lower temperature, convergence is slow and sometimes
unstable or barely stable. In that case taking
\begin{equation}
\sigma_{\alpha,i}=-s_{\alpha}{ \rm
sign}(V_\alpha)\langle\hat{O}_{\alpha}\rangle _{\sigma=0}
\end{equation}
for a starting configuration also leads to a shorter thermalization
time than the choice of $\sigma=0$.

\subsection{Examples}

We now describe several examples to demonstrate our methods for
calculating nuclear properties. Two major considerations arise in the
implementation: the choice of decomposition scheme and the choice of
ensemble. Different decomposition schemes involve different
dimensions of matrices and numbers of fields, which control the
computational speed. Also, the rates of convergence as $\Delta \beta
\rightarrow0$ are different and determine the number of time slices
to be used. Finally, and perhaps most importantly, this choice also
affects the ``sign problem'' associated with the statistical
stability of the calculation. Different decomposition schemes will be
compared in Examples 1 and 2 below.

The choice of ensemble among zero-temperature, canonical, and grand
canonical ensembles usually do not affect the issues noted above.
Instead, it depends on the kind of properties to be calculated. The
zero-temperature formalism with a good trial wavefunction is
effective in projecting out the ground state and is suitable for
calculating ground-state static observables. For calculating finite
temperature properties, the canonical ensemble is physically most
relevant but also is most difficult to implement. Examples 1, 2, and
3 below demonstrate the grand canonical, zero-temperature, and
canonical ensembles, respectively.

Finally, we choose a particular nucleus, ${}^{20}$Ne, to demonstrate
the calculation of the dynamical responses for different operators
and recover the strength function by the MaxEnt Method. The examples
shown below were done with 3000 to 6000 samples generated on the
passively parallel Touchstone Gamma and Delta computers at Caltech.

\begin{center}
{\it Example 1: Monopole pairing interaction in the {\it sd}-shell}
\end{center}

We have described the considerable flexibility in writing the
two-body interaction in quadratic form; e.g., density versus pairing
decomposition and direct versus exchange decomposition. To illustrate
this flexibility, we consider protons only in the {\it sd}-shell
($N_s=12$), and keep only the $J=0$ terms in Eq.~(3.5); the values of
$V_{J=0}$ and single particle energies are taken from the Wildenthal
interaction \cite{wild84}. We first recouple the Hamiltonian into a
quadratic form in the density operators in Eq.~(3.21); because all
possible density operators are required, there are 144 fields for
each time slice. We then use the pairing decomposition in Eq.~(3.41);
after diagonalization only 6 fields are required. Finally, we write
the two-body interaction as $\hat{H}_{2}={1\over 2}\hat{H}_{2}+
{1\over 2}\hat{H}_{2}$ and decompose the first half using densities
and the second half using pairs; the total number of fields in this
case is 150. It turns out that this system is 99\% free of the sign
problem (i.e., $\hbox{Real}(\Phi_i)>0$\ \ 99\% of the time),
independent of the decomposition.

All three calculations were performed in the grand-canonical ensemble
using a Gaussian weight function around the static mean field, at an
inverse temperature of $\beta=1$ (here and henceforth, we measure all
physical energies in MeV) and fixed chemical potential. The
expectation value of the proton number, energy, and $J^{2}$ are given
in Fig.~1 as a function of $\Delta \beta$. As the number of time
slices increases (i.e., $\Delta \beta \rightarrow 0$), all three
decompositions converge to the exact answer, demonstrating their
mutual equivalence in the continuum limit. Note, however, the
different rates of convergence.

\begin{center}
{\it Example 2: ${}^{24}$Mg with schematic forces}
\end{center}

Next we consider {\it sd}-shell nuclei with a schematic Pairing +
Multipole density interaction, where the multipole force is
separable; it is the same interaction used in \cite{johnson92}:
\begin{eqnarray}
\hat{H}_2 & = & -g \sum_{T_z=-1,0,1} P_{T_z}^{\dagger}P_{T_z}-{1\over
2}\chi_0
\bar{\rho}_{0,0} \bar{\rho}_{0,0}\nonumber\\
& & \qquad-{1 \over 2}\chi_2 \sum_M
\bar{\rho}_{2,M}\bar{\rho}_{2,-M}(-1)^M
-{1\over 2} \chi_4 \sum_{M} \bar{\rho}_{4,M} \bar{\rho}_{4,-M}(-1)^M
\end{eqnarray}
where
\begin{equation}
P_{T_z}=\sum_{j m t_1 t_2} ({1\over 2} t_1
{1 \over 2} t_2| 1 T_z)
a_{j m t_1} \tilde{a}_{j m t_2}
\end{equation}
and
\begin{equation}
\bar{\rho}_{K,M}=\sum \tilde{u}_{K}(j_1j_2)
\rho_{K,M,T=0}(j_1,j_2)\;.
\end{equation}
This Hamiltonian was also decomposed in several ways. We first
decomposed the pairing interaction as pair operators and the
multipole force as density operators. In this way, the number of
fields is kept to a minimum (only 21 per time slice). The
pair-operator $P_{T_z=0}$ mixes protons and neutrons and therefore
one matrix representing the mixed neutron and proton wavefunction is
needed. In addition, the pairing decomposition requires matrices
whose dimension is $2 N_s$, so that the matrices involved in this
calculation are $48\times 48$.

We calculated ${}^{24}$Mg (4 valence protons and 4 valence neutrons)
in the zero temperature formalism; i.e., using the evolution operator
at large $\beta$ to project out the ground state from a trial state
$\psi_0$. Since $\hat{h}$ is hermitian (here $\bar{\rho}$ has the
property $\bar{\rho}^{\dagger}_{K,M} = \bar{\rho}_{K,-M}(-1)^{M})$,
in the static path $\zeta=\langle\psi_0|\exp(-\beta
\hat{h})|\psi_0\rangle$ is always positive definite and $\Phi_i = 1$.
However, for calculations with more than one time slice $\langle \Phi
\rangle$ becomes very small, so that we can only obtain sensible
results in the SPA. These results turned out to be extremely good,
relaxing to the right energy and angular momentum (Fig.~2). However,
the success of the static path is case-specific and not well
understood. For example, we have also tried the case of just
multipole interactions among 4 protons in the {\it sd}-shell
(${}^{20}$Mg), and find that the SPA relaxes to an energy 2~MeV
higher than the ground state.

In a second scheme, we transformed the pairing part of the
interaction (6.7) into a quadratic form in the density operators. In
this transformation, we used only density operators that conserve
proton and neutron numbers (3.24--3.34), and chose all $E_{K,T=1}$
elements in Eq.~(3.27) to be zero so that only isoscalar density
operators were present in the quadratic form.

In this case, the interaction is in a much more complicated form due
to the Pandya transformation of the pairing interaction, and one
needs 144 fields per time slice (as compared to 21 fields needed in
the first decomposition). However, an advantage lies in the
separation of the Slater-determinants for protons and neutrons since
only density operators that separately conserve neutron and proton
numbers are present. In the {\it sd}-shell, the dimension of matrix
involved is only $12\times 12$. For this particular interaction, an
even more desirable property of the second scheme is that the
eigenvalues $\lambda_{K,\alpha}$ in Eq.~(3.34) found by diagonalizing
$E_K$ in Eq.~(3.27) satisfy ${\rm sign} (\lambda_K)=(-1)^{K+1}$. We
prove in the next section that this property guarantees $
\zeta(\sigma)$ to be positive definite for even-even nuclei if a
suitable trial state is chosen. This allows calculations with any
number of time slices that are free from any sign problem.

We performed the calculation using the zero-temperature formalism at
different $\beta$ and $\Delta \beta$ values, choosing first the
Hartree Slater-determinant as the trial wavefunction. The SPA
calculation and that with $\Delta \beta =0.125$ are shown in Fig.~2.
We have also performed calculations at $\Delta \beta =0.5$ and
$\Delta \beta=0.25$. These results are not shown but lie between the
SPA and the $\Delta \beta=0.125$ results, and show a convergence to
the result at $\Delta \beta =0.125$. At $\Delta \beta =0.125$, the
energy relaxes to the right energy, whereas the SPA also relaxes, but
to a much higher energy.

We repeated these last calculations with a different trial
wavefunction: a Slater determinant of the orbitals $(j,m)=({5\over
2},\pm {1\over 2}),({5\over 2},\pm {3\over 2})$. Although a different
relaxation curve is traced out by the results at $\Delta \beta
=0.125$, it also converges to the same exact result (Fig.~2). The
choice of the trial state is therefore important for determining the
rate of relaxation of the zero-temperature approach, although the
final result is independent of the trial state as expected. In this
case, although the Hartree state is lower in energy than the maximal
prolate state (compare $\langle H \rangle$ at $\beta =0$), it
contains some high angular momentum components (compare $\langle J^2
\rangle$ at $\beta=0$), so that it relaxes more slowly and reaches
the ground state at a larger value of $\beta$.

\begin{center}
{\it Example 3: Canonical calculations of ${}^{20}$Ne and
${}^{24}$Mg}
\end{center}

Next, we demonstrate the canonical ensemble for the same interaction
(6.7) using the pure density decomposition as described in Example~2.
We calculate ${}^{20}$Ne because it is small enough to allow for an
exact diagonalization to give all the states of $\hat{H}$, since we
are concerned with both ground-state and thermal properties. The
canonical trace for this path-integral in also positive definite (see
Section~VII), allowing the calculations to be done using many time
slices.

The results for calculations with $\Delta \beta =0.25, 0.125$, and
$0.0625$ are shown in Fig.~3. The convergence as a function of
$\Delta \beta$ is also apparent. Note that particular attention
should be given to the extrapolation at high temperature. However, it
is not hard to increase the number of time slices to decrease the
error at high temperature. For example, for $\beta =0.5, \Delta \beta
=0.0625 $ amounts to only 8 time slices. Similar results on
${}^{24}$Mg in the canonical ensemble are shown in Fig.~4. The
relaxation to the ground state can be compared with the
zero-temperature result in Fig.~2; however, now the results at small
values of $\beta$ are also physically significant.

In the calculation of ${}^{20}$Ne, the activity expansion method is
numerically stable. However, instabilities appear for {\it sd}-nuclei
when the number of proton or neutron valence particles (or holes) is
greater than~4. The instability is signalled by the deviation of
$\langle N_p \rangle$ and $\langle N_n \rangle$ from integers. In
those cases, the expansion in Eq.~(4.27) or (4.35) gives the
canonical trace as the small difference between very large numbers,
so that mid-shell nuclei cannot be calculated directly by those
equations. We have developed an alternative Fourier expansion
technique to extract the canonical trace that gives satisfactory
results \cite{dean93}.

The real coherent-state method for the canonical trace for
${}^{24}$Mg gives the same results as the activity expansion.
However, $\langle\Phi_i\rangle$ is not always unity due to the need
to integrate over the coherent states in Eq.~(4.40) (as will be
explained in the next section). It changes from 0.70 to 0.23 for
$\beta$ changing from 0.5 to 1.0 at $\Delta \beta=0.25$.

We have also studied rotating nuclei in the canonical ensemble by
adding a Lagrange multiplier to the Hamiltonian,
$\hat{H}'=\hat{H}-\omega \hat{J}_z$, where $\hat{J}_z$ is the
$z$-component of the angular momentum and $\omega$ is the cranking
frequency. The calculations at $\beta =1$ for ${}^{20}$Ne with
$\Delta \beta=0.125, 0.0625$, and $0.03125$ are shown in Fig.~5,
where the convergence to the exact results can be seen. The moments
of inertia fitted from the three sets of data are 5.10, 5.30, and
4.95, compared to 4.74, the value from the exact curve. By adding the
term linear in $\hat{J}_z$, we break the time reversal symmetry of
$\hat{h}(\sigma)$, which is related to the sign properties of the
weight function $\zeta(\sigma)$. $\langle \Phi\rangle$ decreases from
1 to 0.55 when $\omega$ is increased from 0 to 1.5 at $\Delta \beta
=0.125$ while it decreases from 1 to 0.52 at $\Delta \beta=0.03125$.

\begin{center}
{\it Example 4: Response and strength functions for ${}^{20}$Ne}
\end{center}

Finally, we demonstrate calculation of the imaginary-time response
functions and the reconstruction of the strength functions. The
calculation for ${}^{20}$Ne is done by the activity expansion method
in a pure density decomposition; the Hamiltonian is that of
Eq.~(6.7). The canonical ensemble is more suitable than the zero
temperature formalism for measuring the dynamical response, because
in the latter case many ``boundary'' time slices are needed to
project out the ground state on both the left and the right, and
extra time slices would have to be introduced in the middle to
measure the response. In contrast, the cyclic property of the trace
allows full use to be made of every time slice in the canonical
ensemble. We choose $\beta=2.5$ (the system has already reached the
ground state at this low temperature, as can be seen from Fig.~3) and
calculate the response functions at $\Delta \beta=0.125$ and $\Delta
\beta=0.0625$ for several one-body operators: the isovector- and the
isoscalar-quadrupole operators $\hat{Q}_v= \hat{Q}_p- \hat{Q}_n$,
$\hat{Q}_s= \hat{Q}_p+ \hat{Q}_n$, and the isoscalar and isovector
angular momentum operators $\hat{J}_s= \hat{J}_p+ \hat{J}_n$,
$\hat{J}_v= \hat{J}_p- \hat{J}_n$. We choose this latter $1^+$
operator purely out of convenience; $\hat{J}_s$ is just the total
angular momentum, which we verified to produce a constant response
equal to $\langle \hat{J}^2 \rangle$, which follows from the
rotational invariance of $\hat{H}$.

The response functions are shown in a semi-log plot in Fig.~6(a,c,e).
For these Hermitian operators, $R(\tau)$ is symmetric about
$\tau=\beta/2$, so it is shown only for $\tau\leq\beta/2$. The slope
of the plot is approximately the energy of the dominant strength. The
$\hat{J}_v$ and $\hat{Q}_v$ responses relax more rapidly than does
the $\hat{Q}_s$ response, indicating that the two isovector operators
couple to states with higher excitation energy than does the
$\hat{Q}_s$ operator. Since ${}^{20}$Ne has a $J=0$, $T=0$ ground
state, the states excited by an operator will carry the $J$ and $T$
quantum numbers of the operator. The plots are therefore consistent
with the existence of a low-lying $2^+$ state.

The MaxEnt reconstructions of the most probable strength function for
the different one-body operators are shown in Fig.~6(b,d,e). The
reconstruction is performed for each response function measured at
each $\Delta \beta$ value. The figures show the convergence in
$\Delta \beta$ of the resulting strength functions to the exact
strength function. Note the movement of the peaks towards the exact
position and also the decrease in the widths as $\Delta \beta$
decreases.

The average $n$'th moments of the strength function, defined by
\begin{equation}
M_n\equiv\sum_i\omega^n_if(\omega_i)\;;
\end{equation}
can be found in the Monte Carlo integration over all the possible
distributions of the $f_i$. Their uncertainties can be determined
similarly. The $1^{\rm st}$ moment $\langle M_1 \rangle$ and the
$2^{\rm nd}$ moment $\langle M_2\rangle$ of the strength functions
are listed in Table~1 for different operators and different $\Delta
\beta$. The extrapolated moments ($\Delta \beta \rightarrow 0$) and
the exact results for the ground state transitions are also shown.

The single-particle pick-up and stripping response functions for
different orbitals are given in a semi-log plot in Fig.~7(a,c,e) and
Fig.~8. The strength functions extracted from these responses are
related to the excitation spectrum in the neighboring nucleus with
one additional particle or hole. The stripping and pick-up responses
are the same for protons and neutrons as the ground state of
${}^{20}$Ne is isoscalar, and the final states have the angular
momentum of the single nucleon that is added or removed. We see from
Fig.~7 that both the pick-up and stripping responses for the $j={ 5
\over 2}$ orbital and the pick-up response for the $j={1\over 2}$
orbital converge to the exact results as $\Delta \beta$ becomes
small; the MaxEnt reconstruction of the corresponding strength
functions in Fig.~7(b,d,e) also show a convergence to the exact
results. The extracted moments are listed in Table~1. However, the
responses for the $j={3 \over 2}$ orbital show an anomalous behavior:
they are close to the exact result at $\tau=0$, and then, with a
sudden change in slope, follow the responses of the $j={5\over 2}$
orbital. A possible reason is that angular momentum is not conserved
sample-by-sample in the calculation, but rather only statistically.
The $J={5 \over 2}$ states for ${}^{19}$Ne and ${}^{20}$Ne nuclei are
much lower in energy than the $J={3 \over 2}$ states (because the
$j={5 \over 2} $ orbital is lower than the $j={3 \over 2}$ orbital by
5.6~MeV), so that if a small admixture of the $J= {5 \over 2}$ states
``leaks'' into the intermediate states for the $j={3\over2}$
response, it will dominate with an exponentially growing correlation
function. (The $j={1\over 2} $ orbital is much closer to the $j={5
\over 2}$ orbital in comparison, only 0.8~MeV higher.)

\section{The Sign problem and future developments}

We have seen in the examples above that the sign properties for the
function $G(\sigma)\zeta(\sigma)$ are crucial to the stability of the
calculation, and may frustrate attempts to apply the Monte Carlo path
integral to any general two-body interaction. In general, if we
choose an arbitrary two-body interaction and arbitrarily decompose it
into pair or density operators, $\langle \Phi \rangle$ vanishes
rapidly as $\beta$ increases or as the number of time slices $N_t$
increases at a fixed $\beta$. For example, with the Wildenthal
interaction and for any calculation for $\beta > 1 $ with more than
three time slices, the noise due to the sign completely swamps the
signal. This ``sign-problem'' is well-known in auxiliary-field
path-integral calculations \cite{loh90} and fermion quantum Monte
Carlo more generally. If there is no explicit symmetry to enforce the
positivity of $\zeta_\sigma$, $\langle \Phi \rangle $ decays
exponentially as a function of $\beta$, and the problem is more
severe for smaller $\Delta \beta$, so that it is very difficult to
calculate low-temperature properties.

Only a handful of interacting fermion systems are known to give rise
to positive definite path-integral: the one-dimensional Hubbard
Model, the half-filled Hubbard Model, and the attractive Hubbard
Model at any dimension and filling\cite{linden92}. We will show in
the following that an important class of interactions for the nuclear
shell model has a positive-definite path-integral representation for
even-even nuclei. It arises from the symmetry between time-reversed
single-particle orbitals and may serve as a starting point in
understanding and controlling the `sign problem'.

We first define the `time-reversed' partner of each creation and
annihilation operator to be
\begin{mathletters}
\begin{eqnarray}
\tilde{a}_{j,m} & \equiv & a_{j,\bar{m}}=(-1)^{j+m}a_{j,-m}\\
\tilde{a^{\dagger}}_{j,m} & \equiv & a^{\dagger}_{j,\bar{m}}
=(-1)^{j+m}a_{j,-m}^{\dagger}\;.
\end{eqnarray}
\end{mathletters}
Note that
\begin{equation}
\tilde{\tilde{a}}_{j,m}= -a_{j,m}
\end{equation}
due to the spin-half statistics.

The class of Hamiltonians that give rise to a positive-definite path
integral are of the form
\begin{equation}
\hat{H}= -{1\over 2}\sum_{\alpha} \chi_\alpha
\rho_{\alpha}\tilde{\rho}_{\alpha}
+e_{\alpha}\rho_{\alpha}+e_{\alpha}^{*}\tilde{\rho}_{\alpha}
\end{equation}
where $\chi_\alpha > 0 $, $e_\alpha $ can be generally complex, and
$\rho_{\alpha}$ is a general density operator of the form
\begin{equation}
\rho_{\alpha}=\sum_{ij} C_{ij} a_{i}^{\dagger} a_j \;.
\end{equation}
Note the requirement of a negative coupling constant for the density
operator with its time-reversed partner and the time-reversal
invariant form of the remaining one-body part.

For application to the shell model, we refer to Eq.~(3.17), so that
in the density decomposition,
\begin{equation}
\tilde{\rho}_{KM}(\alpha) = (-1)^{K+M} \rho_{K-M}(\alpha)\;.
\end{equation}
The requirement of a negative coupling constant for the density
operators leads to a sign rule for the $\lambda_{K\alpha}$, namely
\begin{equation}
{\rm Sign}(\lambda_{K \alpha})= (-1)^{K+1}\;.
\end{equation}
As we will show below, the Hamiltonian (7.3) gives rise to a one-body
hamiltonian in the path integral, $\hat{h}_\sigma$, that is symmetric
in time-reversed orbitals. Time-reversed pairs of single-particle
orbitals thus acquire complex-conjugate phases in the propagation,
guaranteeing a positive definite overlap function $\zeta$.

After the HS transformation on Eq.~(7.3), the linearized Hamiltonian
is
\begin{equation}
\hat{h}=-\sum_\alpha \chi_\alpha \left( (\sigma_\alpha + i
\tau_\alpha)
\rho_\alpha+ (\sigma_\alpha -i\tau_\alpha)\tilde{\rho}_\alpha
\right)
+ e_\alpha \rho_\alpha +e^*_\alpha \tilde{\rho}_\alpha\;,
\end{equation}
so that ${\rho}$ and $\tilde{{\rho}}$ couple to complex-conjugate
fields. (If some part of the interaction does not satisfy (7.6), then
there will be terms in $\hat{h}$ that are of the form
$i(\sigma_{\alpha} +i\tau_{\alpha}){\rho}_{\alpha} +
i(\sigma_{\alpha} -i\tau_{\alpha})\tilde{{\rho}}_{\alpha}$, and the
above statement is not true.)

We represent the single-particle wavefunction by a vector of the form
\begin{equation}
\pmatrix{jm\cr j\bar{m}}, m >0\;,
\end{equation}
with states with $m >0$ in the first half of the vector and their
time reversed orbitals in the second half. Due to Eq.~(7.3) and the
fact that time reversed operators are coupled to complex conjugate
fields, the matrix $\bbox{M}_i$ representing $\hat{h}_i$ has the
structure
\begin{equation}
\bbox{M_i}=\pmatrix{\bbox{A_i} & \bbox{B_i} \cr
\bbox{-B_i^{*}} & \bbox{A_i^{*}}\cr}\;,
\end{equation}
and one can easily verify that the total evolution matrix,
\begin{equation}
\bbox{U}= \prod_i \exp(-\bbox{M}_i \Delta \beta) =
\pmatrix{\bbox{P} & \bbox{Q}\cr \bbox{-Q^*} & \bbox{P^*}\cr}\;,
\end{equation}
is of the same form. Here, $\bbox{A,B,P,Q} $ are matrices of
dimension $N_s/2$. One can show that this matrix has pairs of complex
conjugate eigenvalues $\epsilon,\epsilon^*$ with respective
eigenvectors $\pmatrix{u\cr v\cr}$ and $\pmatrix {-v^* \cr u^*\cr}$.
In the case where $\epsilon$ is real, it is doubly degenerate since
the two eigenvectors are distinct.

For the grand-canonical ensemble, the overlap function is given by
\begin{equation}
\zeta=\det\left[\bbox{1} +\pmatrix{\bbox{P} & \bbox{Q}\cr -\bbox{Q}^*
& \bbox{P}^*\cr}\right]
=\prod_i^{N_s/2} (1+\epsilon_i)(1+\epsilon_i^*)>0\;.
\end{equation}
If only particle-type (neutron-proton) conserving density operators
are present in Eq.~(7.3), each type of nucleon is represented by a
separate Slater-determinant having the structure (7.10), and
therefore $\zeta=\zeta_p \times \zeta_n >0,$ since $\zeta_p >0$ and
$\zeta_n >0$.

In the zero-temperature formalism, if the trial wavefunction for an
even number of particles is chosen to consist of time-reversed pairs
of single-particle states,
\begin{equation}
\bbox{\Psi_0} = \pmatrix{ \bbox{a} & \bbox{b}\cr \bbox{-b^*} &
\bbox{a^*}\cr}\;,
\end{equation}
where $\bbox{a,b}$ are matrices with dimension $({N_s\over 2}\times
{N_v \over 2})$, then $\bbox{\Psi_0^\dagger U \Psi_0}$ is a
$N_v\times N_v$ matrix with the structure (7.10), and the overlap
function again satisfies
\begin{equation}
\zeta= \det[\bbox{\Psi_0^\dagger U \Psi_0}] >0\;.
\end{equation}
If only particle-type conserving operators are present, then
time-reversed pairs of trial wavefunctions can be chosen for both
protons and neutrons in an even-even nucleus, giving rise to
$\zeta=\zeta_p \times \zeta_n >0$. Note that while the overlap
function is positive definite in the grand-canonical ensemble for any
chemical potential (and thus any average number of particles), it is
true only for an even (or even-even if involving a type-conserving
decomposition) system in the zero-temperature formalism with a
suitable trial wavefunction.

In the canonical ensemble for $N$ particles, a fixed-number trace is
involved and therefore
\begin{equation}
\zeta\left(\prod_i \exp(-\bbox{M}_{i} \Delta \beta)\right)=
\sum_{i1\neq i2 \neq\ldots i_N} \epsilon_{i1}\epsilon_{i2}\ldots
\epsilon_{i_N}\;.
\end{equation}
We have found empirically that $\zeta$ is positive definite for
even-even systems, although we lack a rigorous proof. A special case
of the Hamiltonian (7.2) exists in which the overlap function is
positive definite also for odd-odd $N=Z$ nuclei. The required
condition is that only isoscalar-density operators are present in
Eq.~(7.2). This leads to a further symmetry that protons and neutrons
couple to the same field in Eq.~(7.7) and therefore the evolution
matrices satisfy $\bbox{U_p}= \bbox{U_n} \equiv \bbox{U}$. In the
zero-temperature formalism, if we choose the trial wavefunction for
proton and neutrons to be time reversed partners of each other, so
that
\begin{equation}
\bbox{\Psi}_{0,p} =\pmatrix {\bbox{a}\cr \bbox{b}\cr},
\bbox{\Psi}_{0,n}=\pmatrix{
\bbox{-b^{*}} \cr \bbox{a^{*}} \cr}
\end{equation}
then
\begin{eqnarray}
\zeta_p & = & \det[\bbox{\Psi}_{0,p}^{\dagger}
\bbox{U} \bbox{\Psi}_{0,p}]\nonumber\\
& = & \det [\bbox{a^{\dagger} P\ a +
a^{\dagger} Q \ b -b^{\dagger} Q^* a
+b^{\dagger} P^* b}]
=\zeta_n^*\;,
\end{eqnarray}
and $\zeta=\zeta_p\times\zeta_n >0$. On the other hand, for the
canonical ensemble, one can prove that Eq.~(7.13) is real. Given that
$\zeta_n=\zeta_p$, $\zeta$ is a square and is therefore positive.

For a system with general form (7.3), if we perform the canonical
trace by integration over real coherent states, then there is an
extra operator, $\exp(X_{ph} a_{p}^{\dagger}a_{h})$, multiplying the
evolution operator. Time-reversed states couple to different real
fields in the extra operator, and a sign problem arises, as seen in
Example~3 above.

Cranking also causes the sign to depart from unity. In cranking, the
Lagrange multiplier term $\omega J_z$ is added to $\hat{h}$ in
Eq.~(7.7), destroying the property that time-reversed operators are
coupled to complex-conjugate numbers (because $\tilde{J}_z =-J_z $).
Notice, however, that cranking with an imaginary Lagrange multiplier
$i \omega J_z$ will preserve the time-reversal symmetry and will give
rise to positive path-integral.

In summary, for a Hamiltonian of the form (7.3), the above proof
guarantees the overlap function to be positive for any nucleus in the
grand canonical ensemble, and for even-even nuclei in either the
canonical ensemble or zero-temperature formalism (with suitable trial
wavefunction). It also guarantees positivity for odd-odd $N=Z$
systems when only isoscalar density operators are involved.

The Hamiltonian (6.7) satisfies (7.6) upon decomposition of the
paring interaction using density operators and involves only
isoscalar operators, so that $\zeta$ is positive for even-even and
$N=Z$ nuclei. For other systems, $\langle \Phi \rangle $ decreases as
a function of $\beta$. At $\Delta \beta=0.0625$, $\langle \Phi
\rangle = 0.4$ at $ \beta=1.5$ for ${}^{24}$Na and $\langle \Phi
\rangle =0.2$ at $\beta=2.0$ for ${}^{23}$Na. Thus, even for odd-A
nuclei, the sign properties of (6.7) are still much better than that
of a general interaction violating the criteria (7.6). For example,
using the Wildenthal interaction, $\langle \Phi \rangle$ drops to
several percent at $\beta=1$ for any nucleus.

Arbitrary two-body interactions do not satisfy the sign rule (7.6),
and when the rule is violated, $\langle \Phi \rangle$ rapidly
decreases as a function of $\beta$. Note that monopole pairing plays
an important role here. The pairing interaction can be written as
\begin{equation}
-g \sum_{ij} a_i^\dagger a_j \tilde{a^\dagger}_i \tilde{a}_j\;.
\end{equation}
It produces a constant shift in every $\chi_\alpha$ in Eq.~(7.3), as
may be seen also from the multipole decomposition of the pairing
force:
\begin{equation}
-g \sum_{\alpha}(-1)^K\rho_{KM}(\alpha) \rho_{K-M}(\alpha) (-1)^M\;.
\end{equation}
Therefore, if the pairing interaction is strong enough compared to
the remaining interactions, the sign rule can be satisfied.

\section{Conclusion}

We have developed a general framework for carrying out
auxiliary-field Monte Carlo calculations of the nuclear shell model.
In this framework we evaluate ground state or thermal observables,
using pairing or density fields or both.

Although the use of pairing fields naturally embodies important
aspects of the residual interaction, these calculations are more
difficult due to the larger matrix dimension needed and the extra
effort to keep track of the sign of the overlap function as the
wavefunction is propagated. Furthermore, for calculations with
multiple time slices, the Monte Carlo method with a pairing
decomposition suffers from severe sign problems. However, pairing
fields are suitable for carrying out static path or two-time-slice
calculations where the linearized Hamiltonian is hermitian, thereby
enforcing the positivity of the overlap function. This can be easily
verified by observing that for hermitian $\hat{h}, \hat{h}_a,$ and
$\hat{h}_b$, with real eigenvalues $E_i, E_{i_a}$ and $E_{i_b}$,
\begin{equation}{\rm \hat{T}r}[e^{-\beta \hat{h}}]=\sum_i e^{-\beta
E_i}>0,
\end{equation}
and
\begin{equation}
{\rm \hat{T}r}[e^{-{\beta \over 2} \hat{h}_a} e^{-{\beta \over 2}
\hat{h}_b}]
=\sum_{i_a} e^{-{\beta \over 2} E_{i_a}} \langle i_a |e^{-{\beta
\over 2}
\hat{h}_b}|i_a \rangle>0\;.
\end{equation}
In these cases, there is no sign problem and also there is no need to
keep track of the evolution of the sign.

For the density decomposition, we have found a class of interactions
which give rise to a positive definite integrand upon the HS
transformation. For these interactions, stable calculations can be
carried out for many time slices to extrapolate to the exact results
($\Delta \beta \rightarrow 0$). This class of interactions includes
the phenomenological pairing-plus-multipole interaction used widely.
We have carried out calculations with such interactions in the {\it
sd}-shell, demonstrating the power of the method in calculating both
static and dynamical properties in the ground state and at finite
temperature; high spin nuclei were also studied by cranking. The
calculations converge to the exact results (as found by direct
diagonalization) with increasing number of time slices. Although the
nuclear wavefunction is not found explicitly in these calculations,
many nuclear properties can be obtained.

For general shell model interactions, it appears that the sign or
phase property of the integrand is the major factor determining
successful application of the Monte Carlo sampling. Successful
calculations are usually confined to high temperature studies. We
have demonstrated the freedom one has in the decomposition scheme of
the two-body interaction and found that the behavior of the sign can
be different in the various schemes. The next crucial step is to
explore whether we can manipulate these degrees of freedom to enable
stable calculations of nuclei using general forces.

\acknowledgements
This work was supported in part by the National Science Foundation,
Grants PHY90-13248 and PHY91-15574, and by a DuBridge Fellowship held
by W.E.O. We are grateful for discussions with P.~Vogel and D.~Dean
and thank B.~Girish for his help with the Intel parallel
supercomputers.

\newpage
\appendix
\section{Derivation of the overlap formulae for pairing fields}

We consider operators of the form
\begin{equation}
\hat{U}=\exp(-\Delta \beta \hat{h}_{N_t} )
\exp(-\Delta \beta \hat{h}_{N_t-1})
\ldots\exp(-\Delta \beta \hat{h}_1)\;,
\end{equation}
where each $\hat{h}_{t}$ is a quadratic operator
\begin{equation}
\hat{h}_{t}=\sum_{i,j=1}^{Ns} \Theta(t)_{ij}
a_{i}^{\dagger} a_{j}
+ \Delta(t)_{ij} a_{i}^{\dagger} a_{j}^{\dagger} +
\Lambda(t)_{ij} a_i a_j\;.
\end{equation}
Without loss of generality, we choose $\Delta, \Lambda$ to be
anti-symmetric ($\Delta=-\Delta^{T}$, etc.). We follow the
development of Berezin \cite{ber66}, who considered the special case
$\hat{U} =\exp(-i\hat{h}t)$ with $\hat{h}$ Hermitian; we take the
general case.

We calculate the grand-canonical trace
\begin{equation}
{\rm \hat{T}r}\hat{U}=\sum_i \langle i | \hat{U} | i \rangle
\end{equation}
(the sum is over all states of all particle number) by using the
fermion coherent state (FCS) representation of unity \cite{negele}
\begin{equation}
\bbox{1}=\int \prod_{\alpha} d \xi_\alpha d \xi^{*}_{\alpha}
\exp(-\sum_\alpha
\xi_\alpha^{*} \xi_\alpha) |\xi\rangle \langle \xi|.
\end{equation}
Here ${\xi_\alpha}$ are Grassman variables and the $|\xi\rangle$ are
fermion coherent states. Then
\begin{equation}
{\rm \hat{T}r}\hat{U}=\int \prod_\alpha d\xi_\alpha d\xi^{*}_\alpha
\exp(-\sum_\alpha \xi_\alpha^{*}\xi_\alpha) \langle \xi | \hat{U} |
\xi
\rangle.
\end{equation}

We need the FCS representation of $\hat{U}$. In what immediately
follows we show that if $\bbox{U}$ is the matrix representation of
$\hat{U}$, that is
\begin{mathletters}
\begin{eqnarray}
\bbox{U} & \equiv & \pmatrix{\bbox{U^{11}} &
\bbox{U^{12}} \cr
\bbox{U^{21}} & \bbox{U^{22}} \cr}\\
& \equiv & \exp(-\bbox{M}(N_t)\Delta \beta)
\exp(-\bbox{M}(N_t-1)\Delta \beta)\ldots
\exp(-\bbox{M}(1) \Delta \beta),
\end{eqnarray}
\end{mathletters}
where
\begin{equation}
\bbox{M}(t)\equiv \pmatrix{\Theta(t) &
2 \Delta(t)\cr 2\Lambda(t) & -
\Theta^T(t)\cr},
\end{equation}
then
\begin{equation}
\langle \xi|\hat{U}|\xi \rangle = C \exp \left({1 \over 2}
\pmatrix{\xi & \xi^{*}\cr}
\pmatrix{\bbox{B^{11}} & \bbox{B^{12}} \cr
\bbox{B^{21}} & \bbox{B^{22}} \cr}
\pmatrix{\xi \cr \xi^*}\right)\;,
\end{equation}
with
\begin{mathletters}
\begin{eqnarray}
\bbox{B^{11}} &=& \bbox{U^{22}}^{-1} \bbox{U^{21}}\;,\quad
\bbox{B^{12}}= -\bbox{U^{22}}^{-1}\;,\\
\bbox{B^{21}} &=& (\bbox{U^{22}}^{-1})^T\;, \quad
\bbox{B^{22}}=\bbox{U^{12}}\bbox{U^{22}}^{-1}\;,
\end{eqnarray}
\end{mathletters}
and
\begin{equation}
C=\det(\bbox{U^{22}})^{1\over 2}\exp(-{{\Delta \beta} \over 2}
\sum_{n=1}^{N_t} {\rm Tr} \Theta(n)).
\end{equation}
In this case the trace becomes a Gaussian integral over Grassman
variables; the result is given beginning with equation (A29) below.
However, to come to that point we must derive equations (A8-A10).

To this end, we employ the standard rules for operating on
$|\xi\rangle, \langle \xi |$ with $a^{\dagger},a$:
\begin{mathletters}
\begin{eqnarray}
\langle \xi|\hat{U}\hat{a_\alpha}|\xi\rangle &=&
\langle \xi |\hat{U}|\xi
\rangle \xi_\alpha\\
\langle \xi | \hat{U} a_{\alpha}^{\dagger}
|\xi\rangle &=&
\langle \xi | \hat{U} |\xi \rangle {\overleftarrow{\partial} \over
{\partial \xi_\alpha}}\\
\langle \xi | a_{\alpha}^{\dagger}\hat{U}|\xi\rangle &=&
\xi_{\alpha}^{*} \langle \xi |\hat{U}|\xi \rangle\\
\langle \xi |a_\alpha \hat{U}|\xi \rangle &=&
{{\partial}\over {\partial \xi_{\alpha}^{*}}}\langle \xi | \hat{U} |
\xi \rangle\;.
\end{eqnarray}
\end{mathletters}
Next, we derive expressions for $a_{\alpha}\hat{U},
a_{\alpha}^{\dagger} \hat{U}$, as linear combinations of $\hat{U}
a_{\alpha}, \hat{U} a_{\alpha}^{\dagger}$. Then, with the ansatz (A8)
for $\langle \xi|\hat{U}|\xi\rangle $ as a Gaussian in the Grassman
variables $\xi, \xi^{*}$, we use (A11) to derive the elements
$\bbox{B}$ of the Gaussian given in (A9).

To do this, we introduce the operators $b$, $\bar{b}$ (which are not
necessarily Hermitian conjugates)
\begin{mathletters}
\begin{equation}
b_{\alpha} \equiv \hat{U}^{-1} a_{\alpha} \hat{U}, \quad
\bar{b}_{\alpha} \equiv \hat{U}^{-1} a^{\dagger} \hat{U}\;.
\end{equation}
Then
\begin{equation}
a_\alpha \hat{U} = \hat{U} b_\alpha,\quad
a^{\dagger}_\alpha\hat{U}=\hat{U} \bar{b}_\alpha\;,
\end{equation}
\end{mathletters}
and we seek $b, \bar{b}$ as linear combinations of $a^{\dagger},a$.

Define
\begin{equation}
a_\alpha(\tau)=e^{\hat{h}\tau} a_{\alpha} e^{-\hat{h}\tau}\;.
\end{equation}
Then
\begin{equation}
{d \over {d\tau}} a_\alpha(\tau)=
[\hat{h}, a_{\alpha}(\tau)]\;,
\end{equation}
and similarly for $a_{\alpha}^{\dagger}(\tau).$
Putting all the $a_{\alpha}(\tau), a_{\alpha}^{\dagger}(\tau)$
into a single vector, and using the representation (A2) for
$\hat{h}$, one
finds
\begin{equation}
{d \over {d \tau}} \pmatrix{a(\tau)\cr a^{\dagger}(\tau)\cr}
= -\bbox{M} \pmatrix{a_(\tau) \cr a^{\dagger}(\tau)\cr}\;,
\end{equation}
with $\bbox{M}$ given by (A7).

Solving the differential equation (A15),
\begin{equation}
\pmatrix{a(\tau)\cr a^{\dagger}(\tau)\cr} =
\exp(-\bbox{M} \tau) \pmatrix{a \cr a^{\dagger}\cr}\;,
\end{equation}
and so in general
\begin{eqnarray}
\pmatrix{b \cr \bar{b}\cr} & = & \exp(\Delta \beta \hat{h}_1) \ldots
\exp(\Delta \beta \hat{h}_{N_t}) \pmatrix{a\cr a^{\dagger}\cr}
\exp(-\Delta\beta \hat{h}_{N_t}) \ldots \exp(-\Delta \beta \hat{h}_1)
\nonumber\\
& = & \exp(-\Delta \beta \bbox{M}_1)\ldots
\exp(-\Delta \beta \bbox{M}_{N_t})
\pmatrix{a\cr a^{\dagger}\cr}
\nonumber\\
& = & \pmatrix{\bbox{U^{11}} & \bbox{U^{12}} \cr
\bbox{U^{21}} & \bbox{U^{22}} \cr}
\pmatrix{a\cr a^{\dagger}\cr}\;.
\end{eqnarray}
Then
\begin{mathletters}
\begin{eqnarray}
b_{\alpha} & = & \bbox{U^{11}}_{\alpha \gamma}a_\gamma +
\bbox{U^{12}}_{\alpha\gamma} a_{\gamma}^{\dagger}\;,\\
\bar{b}_\alpha & = & \bbox{U^{21}}_{\alpha \gamma} a_\gamma +
\bbox{U^{22}}_{\alpha \gamma} a_\gamma^{\dagger}\;,
\end{eqnarray}
\end{mathletters}
where the summation on $\gamma$ is implicit. Upon inserting (A18) in
(A12b), and using the ansatz (A8), one can
straightforwardly derive the $\bbox{B}$'s in terms of the
$\bbox{U}$'s as
given in (A9).

Although we do not show it in detail, we note that $\bbox{B}$ is
antisymmetric
($\bbox{B^{11}}^T= -\bbox{B^{11}}, \bbox{B^{21}}^T= -\bbox{B^{12}}, $
etc.),
which can be proved using
\begin{equation}
\pmatrix{\bbox{0} & \bbox{1}\cr \bbox{1} & \bbox{0}\cr} \bbox{M}
\pmatrix{\bbox{0} & \bbox{1} \cr \bbox{1} & \bbox{0} \cr} =
-\bbox{M}^T
\end{equation}
and
\begin{equation}
\pmatrix{\bbox{0} & \bbox{1}\cr \bbox{1} & \bbox{0} \cr}
\bbox{U} \pmatrix{ \bbox{0} & \bbox{1} \cr \bbox{1} &
\bbox{0}}=(\bbox{U}^{-1})^T.
\end{equation}

Now we must show the normalization $C$ is of the form (A10). To do
so,
we find a differential equation for $C$. Letting
\begin{equation}
\hat{U}_{n+1}=\exp(-\Delta \beta \hat{h}_{n+1})\hat{U}_n\;,
\end{equation}
we define
\begin{equation}
\hat{U}_n(t)=\exp(-t\hat{h}_{n+1})\hat{U}_n
\end{equation}
so $\hat{U}_{n+1}=\hat{U}_n(\Delta \beta)$. Taking the expectation
value of (A22) between FCS's, invoking (A11) and equating the parts
independent of $\xi,\xi^{*}$, one obtains
\begin{equation}
{d \over {d t}}\ln C_n(t)= -{\rm Tr}
\left(\Lambda(n+1)\bbox{B^{22}}_n(t)\right)\;.
\end{equation}
Upon differentiating (A10), one obtains
\begin{equation}
{d\over {d t}} \ln C_n(t)={1 \over 2} {\rm Tr}{d \over {d t}}
\ln \bbox{U^{22}}_n(t)-{1 \over 2} {\rm Tr}\Theta_{n+1}\;,
\end{equation}
Using $\bbox{U}_n(t)=\exp(-\bbox{M}_{n+1}t)\bbox{U}_n$, one derives
\begin{equation}
{d\over {d t}}\bbox{U^{22}}_n = -2 \Lambda_{n+1}\bbox{U^{12}}_n
+\Theta_{n+1}^T\bbox{U^{22}}
\end{equation}
and (A24) becomes (A23). Thus Eq.~(A10) satisfies the differential
equation for $C$, and it just remains to establish the overall
normalization.
This is found by choosing $\bbox{M}=0$, so that
$\hat{U}=1,C=1$, and
\begin{equation}
\langle \xi |\hat{U}|\xi\rangle =\exp(-\sum_\alpha \xi_\alpha^{*}
\xi_\alpha)
\end{equation}
which is $ \langle \xi | \xi \rangle$. Thus we have established the
form (A8) for $\langle \xi |\hat{U}| \xi \rangle$.

The integral (A5) is straightforward (see Berezin\cite{ber66}; the
magnitude is
\begin{equation}
{\rm \hat{T}r}\hat{U}= C
\det\pmatrix{\bbox{B^{11}} & \bbox{B^{12}}-\bbox{1}\cr
\bbox{B^{21}}+\bbox{1} & \bbox{B^{22}}}^{1\over 2}\;.
\end{equation}
The phase of ${\rm \hat{T}r}\hat{U}$, though critical, is more
difficult to
obtain (see Appendix B for details).

One can rewrite (A27) in a more compact form. The constant C contains
the
factor $\det(\bbox{S^{22}})^{1 \over 2}$, which can be written using
\begin{equation}
\det\bbox{U^{22}}=
\det\pmatrix{\bbox{U^{22}} & \bbox{U^{12}} \cr \bbox{0} &
\bbox{1}\cr} =\det\pmatrix{\bbox{U^{12}}^T & \bbox{U^{22}}^T \cr
\bbox{1} & \bbox{0}\cr} (-1)^{{N_s}^2}\;.
\end{equation}
Upon introducing (A28) into (A27), performing some algebra, and using
relationships
from (A20), one arrives at
\begin{equation}
{\rm \hat{T}r}\hat{U} =\det(\bbox{1}+ \bbox{U})^{1\over 2}
\exp(-{{\Delta \beta} \over 2} {\rm Tr}\sum_{n=1}^{N_t}\Theta(n))\;,
\end{equation}
where, again, $\bbox{U}$ is the matrix in (A6) representing the
evolution
operator $\hat{U}$.

As for the density case, one can introduce an activity expansion to
project out
an exact particle number
\begin{equation}
{\rm \hat{T}r}( \lambda^{\hat{N}}\hat{U})
=\det(\bbox{1} + \pmatrix{\bbox{\lambda} & \bbox{0}\cr \bbox{0} &
\bbox{1\over \lambda}\cr}
\bbox{U})^{1\over 2} \exp \left(-{{\Delta \beta} \over 2}
\sum_{n=1}^{N_t}
{\rm Tr}\Theta(n)\right)\lambda^{N_s/2}\;.
\end{equation}
Because the matrices $\bbox{1}$ and $\bbox{U}$ are of dimension $2
N_s \times
2 N_s $, one can write $\lambda^{N_s/2}=\det \pmatrix{\bbox{1} &
\bbox{0} \cr
\bbox{0} & \bbox{\lambda}^{1\over 2}}$ and (A30) becomes
\begin{equation}
{\rm \hat{T}r}\left(\lambda^{\hat{N}}\hat{U}\right)=\det
\left(\pmatrix{\bbox{1} & \bbox{0} \cr \bbox{0} & \bbox{\lambda} \cr}
+ \pmatrix{\bbox{\lambda} & \bbox{0} \cr \bbox{0} & \bbox{1} \cr}
\bbox{U}\right)^{1\over 2} \times \exp \left(-{{\Delta \beta}\over 2}
\sum_{n=1}^{N_t}{\rm Tr} \Theta(n) \right).
\end{equation}
This can be expanded into a polynomial in $\lambda$, which then gives
the
canonical ensemble.

Finally, we give the expectation value of $\langle \psi_t|\hat{U}|
\psi_t \rangle$.First we note that the
vacuum expectation value $\langle 0 |\hat{U}|0 \rangle$ is the term
in
(A31) independent of $\lambda$,
\begin{mathletters}
\begin{eqnarray}
\langle 0 | \hat{U} | 0 \rangle & = & \det \pmatrix{\bbox{1} &
\bbox{0}\cr
\bbox{0} & \bbox{U^{22}}\cr}^{1/2}\exp\left(-{{\Delta \beta}\over 2}
\sum_n {\rm Tr}
\Theta(n) \right)\\
& = & \det\left( \pmatrix{\bbox{0} & \bbox{1}\cr} \bbox{U}
\pmatrix{\bbox{0} \cr
\bbox{1} \cr}\right)^{1\over 2} \exp \left(-{{\Delta \beta} \over 2}
\sum_n {\rm Tr}\Theta(n)\right)\;.
\end{eqnarray}
\end{mathletters}

Any quasi-particle excitations can be represented as
Hartree-Fock-Bogoliubov vacua for properly defined new quasi-particle
operators, which corresponds to doing a similarity transformation on
the matrices. If $ |\psi_t\rangle$ is the vacuum to the
quasi-particle annihilation operator $\beta_i$; i.e.,
\begin{equation}
\beta_i|\psi_t\rangle =0,\quad
\beta_i=\bbox{u}_{ij}a_j+\bbox{v}_{ij}a_j^{\dagger},
\end{equation}
then
\begin{eqnarray}
\langle \psi_t |\hat{U} | \psi_t \rangle & = & \det
\left(
\pmatrix{\bbox{0} & \bbox{1} \cr}
\pmatrix{\bbox{u} & \bbox{v} \cr\bbox{v}^{*} & \bbox{u}^{*}\cr}
\bbox{U}
\pmatrix{\bbox{u}^{\dagger} & \bbox{v}^T\cr\bbox{v}^{\dagger} &
\bbox{u}^T \cr}
\pmatrix{\bbox{0} \cr \bbox{1} \cr}\right)
\times \exp \left(-{{\Delta \beta} \over 2}\sum_n {\rm Tr}\Theta(n)
\right)
\nonumber\\
& = & \det \left(\pmatrix{ \bbox{v}^{*} & \bbox{u}^{*} \cr} \bbox{U}
\pmatrix{\bbox{v}^T\cr \bbox{u}^T \cr}\right)^{1\over 2}
\exp \left( -{{\Delta \beta} \over 2} \sum_n {\rm Tr}
\Theta(n)\right).
\end{eqnarray}

\newpage
\section{Sign of the Overlap for Pairing Fields}

The formulae derived in Appendix A for calculating the overlap
$\zeta_{\sigma}(\beta)=
\langle\psi_{L}|\hat{U}(\beta,0;\sigma)|\psi_{R}\rangle$ in the
zero-temperature formalism and ${\rm
\hat{T}r}[\hat{U}_\sigma(\beta,0)] $ in the thermal formalisms all
involve the square root of a determinant, leaving the phase of
$\zeta_\sigma(\beta)$ undetermined by a factor of $\pm1$. This
ambiguity is irrelevant for the Monte Carlo random walk as we
typically take $|\zeta|$ as the weight function, but the phase must
be known unambiguously for calculation of observables, as important
cancellations may result.

We determine the phase by following the evolution of
$\zeta_\sigma(\tau)$ and its derivatives with respect to $\tau$ as
$\tau$ goes from $0$ to $\beta$. For example, if $\zeta_\sigma(\tau)$
were purely real, zero-crossing (with $\zeta'(\tau) \ne 0$) would
indicate a change in the phase by -1. The initial phase at $\tau=0$
is real and positive. Following the evolution is computationally
expensive, but as most of the time is spent on the random walk, where
the phase is irrelevant, the overall computational time is
negligible.

In what follows we give the formulae for up to fourth derivatives for
each of the different formalisms.

\medskip
\centerline{\bf Grand Canonical Ensemble}

Define
\begin{equation}
f(t)=\zeta_{\sigma}(k\Delta \beta +t)={\rm \hat{T}r}[e^{-h_{k+1}t}
\hat{U}_\sigma(k \Delta \beta,0)],
\end{equation}
then
\begin{equation}
f(t)=\det[1+e^{-\bbox{M}_{k+1}t}
\bbox{U}]^{1 \over 2} \exp(-{{\Delta \beta} \over {2}}
\sum_{i=1}^{k} {\rm Tr}[\Theta_{i}]-{t \over 2} {\rm
Tr}[\Theta_{k+1}])
\end{equation}
\begin{equation}
\ln(f)={1 \over 2}{\rm Tr}[\ln(1+e^{-\bbox{M}_{k+1}t}
\bbox{U})]-{{\Delta \beta}\over {2}}
\sum_{i=1}^{k} {\rm Tr}[\Theta_{i}]-{t \over 2} {\rm
Tr}[\Theta_{k+1}]\;.
\end{equation}

Using the abbreviation $\bbox{M}=\bbox{M}_{k+1},
\Theta=\Theta_{k+1}$, let
\begin{equation}
\bbox{G}=(1+e^{-\bbox{M}t}\bbox{U})^{-1}
e^{-\bbox{M}t} \bbox{U}= 1-(1+e^{-\bbox{M}t}
\bbox{U})^{-1},
\end{equation}
\begin{eqnarray}
{{\partial \bbox{G}} \over {\partial t}} & = &
-[1+e^{-\bbox{M}t}\bbox{U}]^{-1}
\bbox{M}e^{-\bbox{M}t}\bbox{U}
[1+e^{-\bbox{M}t}\bbox{U}]^{-1}
\nonumber\\
& = & -\bbox{(1-G)MG}\;.
\end{eqnarray}

The derivatives of $\ln(f)$ can be expressed in terms of the matrices
$\bbox{G}$
and $\bbox{M}$,
\begin{eqnarray}
g_{1} & \equiv & {{\partial \ln(f)}\over {\partial t}} = -{1 \over 2}
{\rm Tr}[\bbox{MG}]-{1 \over 2}{\rm Tr}[\Theta]\\
g_{2} & \equiv & {{\partial^{2} \ln(f)} \over {\partial t^{2}}}=
{1 \over 2} {\rm Tr}[\bbox{M(1-G)MG}]\\
g_{3} & \equiv & {{\partial^{3} \ln(f)} \over {\partial t^{3}}}=
-{1 \over 2} {\rm Tr}[\bbox{MGM(1-G)M(1-2G)}]\\
g_{4} & \equiv & {{\partial^{4} \ln(f)} \over {\partial t^{4}}}=
-{1 \over 2}{\rm Tr}[\bbox{M(1-G)MGM(1-G)M(1-2G)}]
\nonumber\\
& & -{1 \over 2}{\rm Tr}[\bbox{MGM(1-G)MGM(1-2G)} ]
-{\rm Tr}[\bbox{GM(1-G)M(1-G)MGM}]\;.
\end{eqnarray}

Then $\zeta_{\sigma}(k \Delta \beta +t)$ is given by,
\begin{equation}
\zeta_{\sigma}(k \Delta \beta +t)=f(t)
=f(0) \exp\left(g_1 t+g_2 {{t^2}\over {2!}} + g_3 {{t^3}\over {3!}} +
g_4 {{t^4} \over {4!}}+ \ldots \right)\;.
\end{equation}

\centerline{\bf Zero temperature formalism}

In this case
\begin{eqnarray}
f(t) & = & \zeta_{\sigma}(k \Delta \beta+t)=
\langle\psi_{t}|e^{-\hat{h}_{k+1}}t\hat{U}
(k \Delta \beta,
0;\sigma)|\psi_{t}\rangle=\langle\psi_{L}
|e^{-\hat{h}_{k+1}t}|\psi_R\rangle\;,\\
& = & \det[\bbox{\Psi_{L}} e^{-\bbox{M}_{k+1}t}
\bbox{\Psi_R}]^{1\over 2} e^{-{{\Delta \beta}
\over 2} \sum_{i=1}^{k}{\rm Tr}[\Theta_i]-{t \over 2}
{\rm Tr}[\Theta_{k+1}])}\;,\\
\ln(f) & = & { 1\over 2} {\rm Tr}
[\ln(\bbox{\Psi_{L}} e^{-\bbox{M}_{k+1} t}
\bbox{\Psi_R})]-{{\Delta \beta}
\over 2} \sum_{i=1}^{k}{\rm Tr}
[\Theta_{i}]-{ t\over 2} {\rm Tr}[\Theta_{k+1}]\;.
\end{eqnarray}

Let
\begin{equation}
\bbox{G}=e^{-\bbox{M} t}\bbox{\Psi_R} [\bbox{\Psi_L} e^{-\bbox{M}t}
\bbox{\Psi_R}]^{-1}\bbox{\Psi_L}\;,
\end{equation}

\begin{eqnarray}
{{\partial \bbox{G}} \over {\partial t}} & = &
e^{-\bbox{M}t}\bbox{\Psi_R}
[\bbox{\Psi_L} e^{-\bbox{M}t}
\bbox{\Psi_R} ]^{-1}
\bbox{\Psi_L} \bbox{M} e^{-\bbox{M}t} \bbox{\Psi_R}
[\bbox{\Psi_L} e^{-\bbox{M}t}
\bbox{\Psi_R}]^{-1}
\bbox{\Psi_L} \nonumber\\
& & -\bbox{M}e^{-\bbox{M}t}\bbox{\Psi_R}
[\bbox{\Psi_{L}}e^{-\bbox{M}t}\bbox{\Psi_{R}}]^{-1}
\bbox{\Psi_L} \;,\nonumber\\
& = & \bbox{GMG-MG}= -\bbox{(1-G)MG}\;.
\end{eqnarray}
Then
\begin{equation}
g_{1}={{\partial \ln(f)}\over {\partial t}} = -{1 \over 2} {\rm
Tr}[\bbox{MG}]
-{1 \over 2} {\rm Tr}[\Theta]\;,
\end{equation}
and so on and all formulae are the same as in the grand canonical
ensemble (B6-9), except that the matrix $\bbox{G}$ is now different.
In fact, in both cases $\bbox{G}$ can be shown to be the matrix for
the Green's function; i.e.,
\begin{equation}
\bbox{G}_{ij}=\langle \alpha_{j}^{\dagger} \alpha_{i}\rangle
\end{equation}
where
\begin{mathletters}
\begin{equation}
\alpha_i = a_i,i=1,\ldots,N_s
\end{equation}
\begin{equation}
\alpha_{i+N_{s}} = a_{i}^{\dagger},i=1,\ldots, N_s\;.
\end{equation}
\end{mathletters}

\centerline{\bf Canonical Ensemble}

In Eq.~(4.35), the undetermined sign involves only the vacuum
expectation $\langle0|\hat{U}(\beta,0)|0\rangle$. Once it is
determined, the sign for ${\rm \hat{T}r}[\hat{U}(\beta,0)]$ is known.
We can use the equations for the zero temperature formalism to obtain
the sign of $\langle0|\hat{U}(\beta,0)|0\rangle$.

\newpage
\section {Maximum Entropy Extraction of the Strength Function}

We use the Maximum entropy (MaxEnt) method to reconstruct the
strength function from the response function. Here we give a brief
description of the Classic MaxEnt Method; details can be found in the
paper by Gull \cite{gull89}.

The MaxEnt method is a Bayesian approach for reconstruction of
positive additive images $f$ from noisy data. In our case the image
is the strength function ${f}(\omega)$. The noisy data,
$D_j=d_j+\eta_j$, are the measurements of the response function
$R(\tau)$ at the discrete imaginary times, where $d_j=R(j\Delta
\beta)$ and $\eta_j$ is the noise in the data. In the absence of any
data, the most probable image is chosen to be a default model $m$.
Skilling \cite{skill89} proved that, in that case, the only
consistent choice for the probability of an image $f$ is determined
up to a parameter $\alpha$:
\begin{equation}
pr(f)=\exp(\alpha S (f,m))/Z(\alpha,m)\;,
\end{equation}
where $S$ is the entropy of image $f$ relative to the default model.
If the image is discretized to $f_i$ ($i=1,\ldots,r$), then
\begin{equation}
S(f,m)= \sum_j(f_j -m_j -f_j \ln(f_j/m_j))\;,
\end{equation}
and
\begin{equation}
Z_S(\alpha,m)=\int_0^{\infty}d^r f
\prod f^{-{1\over 2}} \exp(\alpha S(f))
=\int_{-\infty}^{\infty} d^r u~\exp(\alpha S(u^2))\;,
\end{equation}
where the last step follows from a change of variable
$u_i=\sqrt{f_i}$. ($Z_S$ has no relation to the nuclear partition
function.)

In the presence of data, we gain some knowledge about the image.
Assuming a Gaussian distribution of errors in the data $D_i$, the
probability for $f$ is
\begin{equation}
pr(f)=\exp(\alpha S -{1\over 2} \chi^2(f))
{1\over {Z_S}} {1\over {Z_L}}\;,
\end{equation}
where
\begin{equation}
\chi^{2}(f)=\sum(d_i(f)-D_i)(d_j(f)-D_j)G^{-1}_{ij},
\end{equation}
$G_{ij} =\langle \eta_i \eta_j\rangle$ is the correlation matrix of
the errors in the data, and
\begin{equation}
Z_L=\int d^{N} D \exp(-{1\over 2} \chi^2(f))\;.
\end{equation}

For a given choice of $\alpha$, the most probable image can be found
by maximizing $\alpha S-{1\over 2} \chi^2$, giving rise to the term
``Maximum Entropy'' method. However, $\alpha$ is not predetermined.
Rather, it is varied until $\chi^2$ at the maximum of $\alpha S
-{1\over 2} \chi^2$ is approximately equal to the total number of
data $D_i$. But this assignment of $\alpha$ is ad hoc and, according
to Gull \cite{gull89}, usually leads to an underfitting of the data.
In the classic MaxEnt, $\alpha$ is fixed by maximizing the
probability of $\alpha$ given the data set $D_i $ and the default
model $m$:
\begin{equation}
pr(\alpha|D,m) \propto Z_Q Z_{S}^{-1}Z_{L}^{-1},
\end{equation}
where
\begin{equation}
Z_{Q} =\int_{0}^{\infty} d^r f \prod f^{1 \over 2} \exp(\alpha S
-{1 \over 2} \chi^2) =\int_{-\infty}^{\infty}d^r u~
\exp(\alpha S-{1\over 2}
\chi^2)\;.
\end{equation}

After $\alpha$ is fixed, we can find the most probable $f$ by
maximizing $\alpha S -{1\over 2}\chi^2$, or we can find the average
$f$ by Monte Carlo sampling using the integrand of (C4) as weight
function. Information about the uncertainty in $f$ can also be
obtained from Monte Carlo sampling. In the particular application at
hand, we would like to know, for example, the uncertainty in the
location of the peaks in $f$, or in the moments $M_n=\int
\tilde{f}(\omega)\omega^n d \omega /\int \tilde{f}(\omega)d\omega$.

Let us now return to the problem of extracting the strength function.
Since the imaginary time response function $R(\tau)$ is given at
discrete times, we can only allow a limited amount of parameters in
the strength function $\tilde{f}$, which we do by discretizing
$\tilde{f}(\omega)$ to $f_i$ at $\omega_i$. To allow for some
smoothness, we choose a Gaussian function centered around each
$\omega_i$, rather than a delta function. Given the imaginary time
response function, we bound the range of $\omega$ over which the $f$
is significant by $\omega_{min}$ and $\omega_{max}$, and choose the
$\omega_i$ to be evenly distributed between them. The number of
$f_n$, $ n_\omega$, should not exceed the number of data $D_i$ we
have, which is the total number of time slices, $Nt$. We choose the
width of each Gaussian, $\Delta \omega$, to be half of the spacing
between the $\omega_i$'s. For a non-hermitian operator $\hat{O}$, the
strength function ${f}(\omega)$ is related to $f_i$ by
\begin{equation}
{f}(\omega)=\sum_i f_i
\exp(-{1\over 2} (\omega-\omega_i)^2/\Delta^2 \omega)
{1\over\Delta\omega\sqrt{2\pi}}\;,
\end{equation}
while for an hermitian operator, ${f}(-\omega)= e^{-\beta
\omega}{f}(\omega)$ in the canonical ensemble, and we choose
\begin{equation}
f(\omega)=\sum_i f_i e^{\beta/2(\omega-\omega_i)}
\left[e^{-1/2(\omega-\omega_i)^2/\Delta\omega^2}+
e^{-1/2(\omega+\omega_i)^2/\Delta\omega^2}\right]
{1\over\Delta\omega\sqrt{2\pi}}\;.
\end{equation}

The response $R(\tau)$ generated by ${f}(\omega) $ is, for
nonhermitian operators,
\begin{equation}
R(\tau)=\sum_i f_i
e^{(-\omega_i \tau+{1\over 2} \tau^2 \Delta \omega^2)}\;,
\end{equation}
while for hermitian operators it is
\begin{equation}
R(\tau)=\sum_i f_i(e^{(-\tau \omega_i)}+
e^{(-(\beta -\tau)\omega_i))}e^{({1\over 2} ({\beta \over 2} -\tau)^2
\Delta \omega^2)}\;.
\end{equation}

We choose the default model $m_i$, $i=1,\ldots,n_\omega $ to be a
constant fixed by $R(0)$, which is related to the total strength. The
data $D_j$ are the values of $R(\tau)$ for $\tau=j \Delta \beta$
measured from the Monte Carlo sampling. The error correlation
function can also be measured in these calculations.

To maximize the probability (C7), we have to know the dependence of
$Z_S$ and $Z_Q$ on $\alpha$. Some simplification can be had by
calculating $Z_S$ in the saddle point approximation. There is a
saddle point in $S(u^2,m)$ at $u_{i}^{2}=m_i$ with the second
derivatives ${{\partial S}\over {\partial u_i \partial
u_j}}|_{u_{i}^{2}=m_i} = -4 \delta_{ij}$, which leads to the
approximate integral
\begin{equation}
\int_{-\infty}^{\infty} d^r u~exp(\alpha S) \simeq
\int_{-\infty}^{\infty}d^r u~exp(-2\alpha \sum_i u_{i}^{2})=
\left({\pi \over {2 \alpha}}\right)^{r/2}\;.
\end{equation}
The condition ${{\partial p} \over {\partial \alpha}} = 0$ then
becomes
\begin{equation}
{1 \over {Z_Q}}\int_{-\infty}^{\infty}d^r u \ S \exp(\alpha S
-{1\over 2}
\chi^2) ={1 \over Z_S}{{\partial Z_S} \over {\partial \alpha}}
= -{r \over {2\alpha}}
\end{equation}
and the average image $\langle f_i\rangle$ is given by
\begin{equation}
\langle f_i\rangle=
{1 \over {Z_Q}}\int_{-\infty}^{\infty}d^r u \ f_i \exp(\alpha S
-{1\over 2}\chi^2)\;.
\end{equation}
We do these integrals by Monte Carlo sampling of $u_{i}$ with
$\exp(\alpha S -{1\over 2}\chi^2) $ as the weight function. The value
$\alpha $ is determined by the self-consistent condition $\langle
S\rangle_\alpha= -{r\over {2 \alpha}}$. When $\alpha$ is known, the
average distribution $\langle f_i\rangle = \langle u_{i}^{2}\rangle$
and the uncertainty $\delta f_i= \sqrt{\langle f_i^2\rangle-\langle
f_i\rangle^2} $ is also given in the course of the Monte Carlo
evaluation of (C15).

\newpage
\begin{figure}
\caption{Calculations in the grand canonical ensemble for protons
only in the {\it sd} shell with monopole interaction (all six
$E_{J=0}$ matrix elements of the Wildenthal interaction), at $\langle
N_p \rangle=3.17$, $\beta =1$. Shown are $\langle H\rangle$ and
$\langle J^2\rangle$ as functions of $\Delta \beta$ for three
different decompositions: pure pairing decomposition, pure density
decomposition, and a half-density and half-pairing decomposition.
Solid diamonds at $\Delta \beta=0$ are the exact results obtained by
direct diagonalization.}
\end{figure}

\begin{figure}
\caption{Zero-temperature calculations of ${}^{24}$Mg with the
schematic interaction (6.7). Note the relaxation of $\langle
H\rangle$ and $\langle J^2\rangle$ as $\beta$ increases. Hollow
triangles are static path calculations in the pure density
decomposition, solid diamonds are static path calculations by
decomposing the pairing interaction into pair operators and the
multipole interaction into density operators. Solid circles and
hollow squares are both calculations in a pure density decomposition
with $\Delta \beta =0.125$, using the Hartree solution and the
maximal prolate state, respectively, as the trial wavefunction. The
solid line segments indicate the exact ground state results.}
\end{figure}

\begin{figure}
\caption{Canonical ensemble calculations of ${}^{20}$Ne with the
schematic interaction (6.7) at $\Delta \beta =0.25, 0.125$ and
$0.0625$ and the exact results; $\langle H\rangle$ and $\langle
J^2\rangle$ are shown as functions of $\beta$. These calculations
were done in a pure density decomposition.}
\end{figure}

\begin{figure}
\caption{Similar to Fig. 3 for ${}^{24}$Mg. }
\end{figure}

\begin{figure}
\caption{Finite temperature cranked calculations of ${}^{20}$Ne with
the schematic interaction (6.7) in the canonical ensemble using a
pure density decomposition. Here $\beta=1 $, with $\Delta \beta
=0.125, 0.0625$ and $0.03125$. The exact cranking curve is also
shown.}
\end{figure}

\begin{figure}
\caption{Canonical ensemble calculations of the response functions
for ${}^{20}$Ne ($\beta=2.5$) at discrete imaginary time using
$\Delta \beta =0.125, 0.0625$, in a pure density decomposition. The
exact results are calculated in the ground state. (a), (c), (e) show
the isoscalar quadrupole ($Q=Q_p+Q_n$), isovector quadrupole
($Q_v=Q_p-Q_n$) and the isovector angular angular momentum
($J_v=J_p-J_n$) responses. The corresponding most probable strength
functions recovered by the MaxEnt method are shown in (b), (d), (f)
respectively. The exact strength functions calculated from ground
state are plotted as discrete lines with the height indicating the
integrated strength of the delta-functions.}
\end{figure}

\begin{figure}
\caption{Similar to Fig. 6. but for the single-particle pickup and
stripping response. (a), (c), (e) show the imaginary time stripping
response for the $j={5 \over 2}$ orbital, and the pickup responses of
the $j={5\over 2}$ and $j={1\over 2}$ orbitals respectively. The
corresponding most probable strength functions recovered by the
MaxEnt method are shown in (b), (d), (f) respectively. The exact
response and strength functions are calculated for the ground state.}
\end{figure}

\begin{figure}
\caption{Similar to Figs. 6,7 but showing the imaginary time pickup
and stripping responses of $j={3\over 2} $ orbital. The response
functions are in agreement with the exact curve for small $\tau$, but
then abruptly follow the $j={5\over 2}$ response at larger $\tau$.}
\end{figure}

\newpage
\begin{table}
\caption{MaxEnt extraction of the moments of the strength functions
corresponding to Figs.~6 and 7. The extrapolated
($\Delta\beta\rightarrow0$) total strength and first two moments are
compared with the exact results for the ground state of ${}^{20}$Ne.}
\begin{tabular}{lccccc}
& & $\Delta\beta=0.125$ & $\Delta\beta=0.0625$ &
extrap & exact\\ \tableline
& total strength & $27.3\pm0.2$ & $25.9\pm0.1$
& $24.5$ & $25.1$\\
$Q(\tau)\cdot Q(0)$ & $\langle\omega\rangle$ &
$2.33\pm0.08$ & $2.77\pm0.08$ & $3.22$ & $3.46$\\
& $\langle\omega^2\rangle$ & $8.09\pm1.2$ &
$10.5\pm1.2$ & $12.9$ & $15.4$\\ \tableline
& total strength & $6.26\pm0.03$ & $6.78\pm0.02$ &
$7.29$ & $6.96$\\
$Q_v(\tau)\cdot Q_v(0)$ & $\langle\omega\rangle$ &
$7.24\pm0.15$ & $7.77\pm0.10$ & $8.31$ & $8.38$\\
& $\langle\omega^2\rangle$ & $59.9\pm3.9$ &
$66.6\pm2.5$ & $73.4$ & $73.8$\\ \tableline
& total strength & $16.3\pm0.1$ & $16.05\pm0.08$ &
$15.8$ & $15.9$\\
$J_v(\tau)\cdot J_v(0)$ & $\langle\omega\rangle$ &
$8.49\pm0.25$ & $9.44\pm0.19$ & $10.39$ & $10.39$\\
& $\langle\omega^2\rangle$ & $89.8\pm9.04$ &
$107.7\pm6.4$ & $125.6$ & $119.6$\\ \tableline
& total strength & $1.59\pm0.01$ & $1.62\pm0.07$ &
$1.64$ & $1.59$\\
$\sum_ma^\dagger_{5/2m}(\tau)a_{5/2m}(0)$ &
$\langle\omega\rangle$ & $9.84\pm0.12$ &
$10.32\pm0.09$ & $10.80$ & $10.98$\\
& $\langle\omega^2\rangle$ & $98.0\pm2$ &
$107.5\pm1.4$ & $117$ & $121$\\ \tableline
& total strength & $4.47\pm0.01$ & $4.42\pm0.09$ &
$4.37$ & $4.41$\\
$\sum_ma_{5/2m}(\tau)a^\dagger_{5/2m}(0)$ &
$\langle\omega\rangle$ & $-3.15\pm0.02$ &
$-3.00\pm0.02$ & $-2.86$ & $-2.81$\\
& $\langle\omega^2\rangle$ & $10.28\pm0.05$ &
$9.71\pm0.04$ & $9.14$ & $10.08$\\ \tableline
& total strength & $1.702\pm0.004$ &
$1.745\pm0.003$ & $1.788$ & $1.773$\\
$\sum_ma_{1/2m}(\tau)a^\dagger_{1/2m}(0)$ &
$\langle\omega\rangle$ & $-3.19\pm0.01$ &
$-3.22\pm0.02$ & $-3.25$ & $-3.16$\\
& $\langle\omega^2\rangle$ & $10.43\pm0.06$ &
$10.65\pm0.04$ & $10.87$ & $11.62$\\
\tableline
\end{tabular}
\end{table}

\end{document}